\documentclass{elsart}
\usepackage{latexsym}
\usepackage[dvips]{graphicx}

\usepackage{epsfig}
\usepackage{changebar}
\usepackage{amssymb}
\usepackage{amsmath}
\usepackage{amsbsy}
\usepackage{subeqnarray}
\usepackage{citesort}

\newcommand{\mc}{\multicolumn}
\def\Tr{{\rm Tr}}
\def\intq#1{\int \frac{d^4#1}{(2\pi)^4}}
\def\intt#1{\int \frac{d^3#1}{(2\pi)^3}}
                                                                                               
\begin{document}
                                                                                               
\begin{frontmatter}

\title{Static properties of nuclear matter within the Boson Loop Expansion}

\author{W. M. Alberico$^\dagger$, R. Cenni$^{\ddagger,\star}$, 
  G. Garbarino$^{\dagger,\ddagger}$ and M. R. Quaglia$^\dagger$}

\address{$\dagger$ Dipartimento di Fisica Teorica, Universit\`a di Torino
  and INFN, \\Sezione di Torino, I--10125 Torino, Italy\\
  $\ddagger$ INFN, Sezione di Genova, I--16146 Genova, Italy \\
  $\star$ Dipartimento di Fisica, Universit\`a di Genova, I--16146 Genova, 
  Italy}

\date{\today}

\begin{abstract}
  The use of the Boson Loop Expansion is proposed for investigating
  the static properties of nuclear matter. We explicitly consider a schematic
  dynamical model in which nucleons interact with the scalar--isoscalar 
  $\sigma$ meson. The suggested approximation scheme is examined in detail 
  at the mean field level and at the one-- and two--loop orders.
  The relevant formulas are provided to derive the binding energy per nucleon,
  the pressure and the compressibility of nuclear matter. 
  Numerical results of the binding energy at the one--loop order
  are presented for Walecka's $\sigma$-$\omega$ model in order to discuss
  the degree of convergence of the Boson Loop Expansion.
\end{abstract}


\end{frontmatter}

\section{Introduction}
\label{sec:1}

Functional integral methods have been extensively used over the years
to provide useful approximation schemes
in many different areas of physics, in both the perturbative and the
non--perturbative regimes~\cite{Co-74,Il-75,Ne-82,It-Zu-80,Amit,NeOr-88}. 
The nuclear many--body problem has been widely investigated 
starting from renormalizable, relativistic Lagrangians based on the so--called
quantum hadrodynamics (QHD)~\cite{SeWa-86}: calculation schemes have been explored,
which go beyond the tree--level approximation and thus involve the dynamics of 
the quantum vacuum. Being a strong--coupling model, no obvious
asymptotic expansion exists for QHD. Hence it is not guaranteed that
a reliable, convergent expansion, allowing for systematic refinements of the 
theoretical predictions, exists for this model. 

Some approaches are based on the so--called Loop Expansion, which can 
be derived from the exact path integral formulation of QHD~\cite{It-Zu-80}: it 
is obtained by expanding the QHD effective action around 
its classical value and by considering terms with increasing number of 
quantum loops~\cite{Co-74,Il-75}. Formally, the resulting Loop Expansion is 
thus a power series in $\hbar$, though Planck's constant is merely a bookkeeping 
parameter~\cite{Il-75}. One can also consider it as a perturbative expansion in 
powers of the coupling constants $g_\sigma$, $g_\omega$, etc, but keeping in mind 
that it is non--perturbative in the mean fields, since it involves Hartree--dressed 
nucleon propagators. The one--loop approximation of the Loop
Expansion corresponds to the Relativistic Hartree Approximation~\cite{chin}.
In Refs.~\cite{Furnstahl:1989wp} full two--loop corrections were 
evaluated for QHD-I (i.e., Walecka's $\sigma$-$\omega$ model) and criteria 
for ``strong'' and ``weak'' convergence were discussed.
Two--loop terms involving the vector meson $\omega$ were found to be
too attractive for standard values of the coupling constant $g_\omega$,
thus making two--loop corrections to the nuclear matter binding energy
too large with respect to the well known mean field results.
This result precluded strong convergence. After tuning the
QHD-I parameters, the correct saturation point of nuclear matter at the
two--loop order was obtained at the price of drastically reducing $g_\omega$.
At the same time $g_\sigma$ was significantly increased, the dynamics of the
system being led almost entirely by the scalar $\sigma$ meson. It was thus 
concluded that, at two--loop order, the Loop Expansion is not even weakly 
convergent.

To suppress the unwanted high--momentum structure of the 
hadronic vacuum loops, short--range correlations and vertex corrections
have also been considered in calculations at the two--loop order 
\cite{vertex1,vertex2,vertex3}. These investigations showed a clear improvement 
over the results of Ref.~\cite{Furnstahl:1989wp}, which were carried out 
with structureless nucleons. Although obtained with a ``reasonable" fit of the 
coupling constants, the results turned out to have a 
strong dependence on the model, not phenomenologically constrained, 
adopted for the short--range correlations or the vertex corrections.
Moreover, the renormalizability of the theory was lost.

The above failure of the Loop Expansion could be due to its inability in
describing the quantum vacuum dynamics. Indeed it is derived as a perturbative 
expansion in the ``large'' QHD-I couplings, while ``small'' couplings are required
to assure some convergence of the Loop Expansion predictions~\cite{LeWi-74}.
One could also claim that the QHD model itself, based on point--like hadrons, 
is not appropriate to include quantum vacuum effects. 

In order to test the validity of these hypotheses,
alternative loop expansions based on QHD, non--perturbative in both 
the couplings and the mean fields, should be studied. In this context 
we mention here the so--called Modified Loop Expansion  or 
Boson Loop Expansion, which is an example of a non--perturbative approach
and in addition preserves renormalizability\footnote{We recall that another 
approach has been recently suggested to overcome the difficulties inherent 
with the description of the vacuum dynamics in QHD: it amounts to utilize
QHD as a non--renormalizable, Effective Field Theory in calculations at 
the tree--level~\cite{Fur-00}. }.

The Modified Loop Expansion, first introduced by Weiss~\cite{Weiss}, 
has been later employed in Ref.~\cite{We90}
together with a chirally symmetric linear $\sigma$ model for nuclear matter 
calculations. In this approach,
fermionic degrees of freedom are integrated out in the generating functional.
This leads to an effective action in which meson loops are included order by
order in $\hbar$ and, at each order $\hbar^n$ in the meson loops,
baryon loops are summed to all orders in $\hbar$.
The zeroth--order of this expansion corresponds to the Relativistic Hartree Approximation,
the modified one--loop level to the Relativistic Random Phase Approximation.
Hence the first--order Modified Loop Expansion corresponds to the two--loop level of the 
Loop Expansion, but with meson propagators dressed in Random Phase Approximation. 
We are not aware of detailed calculations applying this scheme to QHD-I.

The so--called Boson Loop Expansion (BLE), which we shall consider in the present 
paper, has been proposed as a suitable tool to formulate consistent approximation 
schemes for non--perturbative calculations in different issues of nuclear physics
\cite{AlCeMoSa-87,CeSa-88,AlCeMoSa-90,CeSa-94,CeCoSa-97,AmCeSa-99,Al-00}. 
The underlying idea, as in MLE, amounts to integrate out the nucleonic degrees 
of freedom within a path--integral formulation for the dynamics of interacting 
nucleons and mesons. The loop expansion is then carried out on the resulting 
bosonic effective action. 
This approach has been applied to the calculation of the electromagnetic nuclear
response functions and provided a suitable framework to understand the reaction
mechanisms~\cite{CeCoSa-97}. It has also been successfully used 
for the evaluation of the hypernuclear weak decay rates~\cite{Al-00}.

Actually, while the BLE has the merit of rigorously preserving sum rules
and general field theory theorems, the underlying dynamics still needs 
to be more firmly settled. We intend to apply this method to the investigation
of  other observables, such as the static properties of nuclear matter and neutron 
stars, topics of considerable interest and yet under debate after decades of 
valuable works. It is worth noticing that this approach is particularly suited to 
deal with a relativistic description and, in addition, it is stable against sizable 
increases of the nuclear density.

Particular attention has to be payed to the adopted dynamical model. In
a relativistic frame, QHD models are often employed at face to
calculations based on realistic nucleon--nucleon potentials like
the Bonn meson exchange model~\cite{MaHoEl-87}. These two schemes
started from quite different points of view and originally provided sizably 
different results: however in recent times the evolution of calculations in both 
approaches has significantly reduced the gap. QHD is mainly
based on the $\sigma$-$\omega$ interference, while the Bonn
potential, in addition to the exchange of several physical mesons, replaces 
the $\sigma$ meson exchange with a two--pion exchange together 
with the simultaneous excitation of one or two intermediate nucleons to 
$\Delta$ resonances (box diagrams). From a phenomenological point of view,
this seems to favor the Bonn approach for the calculation of
the equation of state of nuclear matter. However, a $G$--matrix calculation 
based on the Bonn potential is quite cumbersome and some self--consistency 
effect of relativistic origin is generally neglected. 
In this context, calculations are usually limited to the Lowest Order Br\"uckner 
Theory, which implies lowest order in the nuclear density; this limitation 
is acceptable in ordinary nuclear matter calculations, but less so in
addressing dense baryonic matter such as in neutron stars.

In order to clarify the fundamental problems underlying the BLE (namely the issue
concerning the renormalization procedure), in this paper we develop the formalism 
within a simple dynamical model containing nucleons and $\sigma$ mesons only.
We shall provide the relevant formulas which are needed to derive the static 
properties of nuclear matter up to the two--boson--loop level. 
In view of a future, comprehensive application to QHD, as well as to give a first 
insight on the convergence properties of the proposed expansion, here we present 
numerical results for the nuclear matter binding energy at the one--loop order
when adopting a $\sigma$-$\omega$ model. The complete BLE formalism for
QHD-I, together with systematic numerical calculations up 
to the two--loop order, will be presented in a forthcoming paper,
also devoted to discuss the phenomenological aspects of the present approach.

The paper is organized as follows. Section \ref{sec:2} contains 
a brief introduction to the general formalism employed in the
functional approach for the evaluation of the static properties
of nuclear matter. In Section \ref{sec:3} a bosonic effective action
is derived for a system of interacting nucleons and $\sigma$ mesons.
The ``elementary" vertices entering into this effective action
are analyzed in some detail in Section \ref{sec:4x}.
In Section \ref{sec:2emezzo} the renormalization procedure
for the bosonic action is carried out.
In Section \ref{sec:4a} we deduce the BLE as a Semiclassical expansion
around the (mean field) solution given by the stationary phase approximation.
Some numerical results for the $\sigma$-$\omega$ model are presented
in Section \ref{numerics}. Finally, in Section \ref{sec:6X} we comment on the
future perspectives opened by the present work.

\section{The functional method -- General formalism}
\label{sec:2}

To construct the formalism we consider a simplified situation 
where a scalar--isoscalar meson (the $\sigma$)  interacts with a massive 
fermion field (the nucleon).
The introduction of the $\omega$ meson, to complete the standard scheme
of QHD induces a mixing with the $\sigma$ meson and a problem of 
renormalizability, 
which can be solved by means of the so--called Stueckelberg gauge \cite{RRA04}
and not simply invoking the current conservation, 
as usually stated \cite{SeWa-86}; this will 
be discussed in a forthcoming paper. 
We shall see in the following that already the ``simple'' scalar field needs
a careful renormalization procedure.

The most general Lagrangian density reads
\begin{equation}
  \label{eq:H001}
  \begin{split}
    {\mathcal L}&=\overline{\psi}(i\rlap/\partial-m)\psi
    +\frac{1}{2}\partial_\mu\sigma\, \partial^\mu \sigma
    -\frac{1}{2}m_\sigma^2\sigma^2+g\overline{\psi}\sigma\psi
    -U(\sigma)\\
    &= \int d^4y\, \overline{\psi}(x)S^{-1}_0(x-y)\, \psi(y)
    +\frac{1}{2}\int d^4y\, \sigma(x) D^{-1}_0(x-y)\, \sigma(y) \\
    &+g\, \overline{\psi}(x)\sigma(x)\psi(x) -U[\sigma(x)]\,,
  \end{split}
\end{equation}
the two expressions differing by an irrelevant four--divergence. In the above
\begin{eqnarray}
S^{-1}_0 (x-y) &=& (i\rlap/\partial_x-m)\, \delta^4(x-y)~, 
\label{Nprop}\\
D^{-1}_0 (x-y) &=& -(\Box_x+m^2_\sigma)\, \delta^4(x-y)~,
\label{sigmaprop}
\end{eqnarray}
formally define the free nucleon and $\sigma$ propagators, $S_0$ and $D_0$ 
respectively, in coordinate space,
and
\begin{equation}
  \label{eq:H222]}
  U(\sigma)=\frac{\lambda_3}{3!}\sigma^3+
  \frac{\lambda_4}{4!}\sigma^4 =
\frac{b\,m\,g^3}{3}\sigma^3+\frac{c\,g^4}{4}\sigma^4\,.
\end{equation}

In momentum space the nucleon propagator in nuclear matter with Fermi
momentum $k_F$ and the free $\sigma$ propagator read, respectively:
\begin{equation}
  \label{eq:019}
  S_0(k)=\frac{\rlap/k+m}{2E_k}
  \left\{\frac{\theta(|{\bf k}|-k_F)}{k_0-E_k+i\eta}+
    \frac{\theta(k_F-|{\bf k}|)}{k_0-E_k-i\eta}
      -\frac{1}{k_0+E_k-i\eta}\right\}~,
\end{equation}
\begin{equation}
  \label{eq:A008}
  D_0(q)=\frac{1}{q^2-m_\sigma^2+i\epsilon}~,
\end{equation}
where $E_k=\sqrt{{\bf k}^2+m^2}$.
It is  also useful to introduce a symbol for the free fermion propagator 
in the vacuum, $S_{0v}\equiv \left. S_0\right|_{k_F= 0}$, 
thus splitting Eq.~(\ref{eq:019}) as follows
\begin{eqnarray}
  \label{eq:H751}
  S_0(k)&=&S_{0v}(k)+S_{0M}(k)~,\\
  S_{0v}(k)&=&\frac{1}{\rlap/k-m+i\eta}~, \\
  S_{0M}(k)&=&(\rlap/k+m) \frac{i\pi}{E_k}\theta(k_F-|{\bf k}|)\delta(k_0-E_k)~,
\end{eqnarray}
the latter being the density--dependent part.

The classical action corresponding to the Lagrangian (\ref{eq:H001}) is given by:
\begin{equation}
  \label{eq:A001}
  {\mathcal A}[\bar\psi,\psi,\sigma]=\int d^4x\, 
  {\mathcal L}(\bar\psi(x),\psi(x),\sigma(x))~,
\end{equation}
where the $\sigma$ and nucleon
fields have to be interpreted as scalar and Grassmann variables respectively.
In a path integral approach the whole dynamics
is deduced from the generating functional
\begin{equation}
  \label{eq:013}
  Z[\bar\eta,\eta,J]=\frac{1}{{\mathcal N}}
  \int D[\bar\psi,\psi,\sigma]\,
        e^{\,i\left\{{\mathcal A}[\bar\psi,\psi,\sigma]+\int d^4x\left[
        \bar\psi(x)\eta(x)+\bar\eta(x)\psi(x)+\sigma(x)J(x)\right]\right\}}~,
\end{equation}
where $\mathcal N$ is a normalization constant while $\eta$, $\bar\eta$ and
$J$ denote the external (classical) sources of the nucleon and $\sigma$
meson fields.
From  $Z$ one  derives the nucleon Green's functions
by functional differentiations with respect to $\eta$ and $\bar\eta$.
The $\sigma$ Green's function is obtained
by derivatives with respect to $J$. 
Other external sources coupled to composite fields could
also be added: for instance, couplings of the kind $\bar \psi(x)\psi(x)\rho(x)$
permit the evaluation of the response function through a double derivative of
$Z$ with respect to $\rho$. 

The static properties of the system follow 
from the partition function $Z(\beta)$
which can be obtained from the generating functional of Eq.~\eqref{eq:013} by:
\begin{enumerate}
\item setting the external sources to 0;
\item performing a Wick rotation, namely
replacing, in the exponent, the time integration
(extended from $-\infty$ to $\infty$) with an integration over the
imaginary time $\tau=it$ in the interval $[0,\beta]$ with
$\beta\equiv \hbar/(k_BT)=1/T$;
\item replacing the Hamiltonian  $\hat H=\int d^3x\,{\mathcal H}(x)$ 
with $\hat H-\mu \hat N$, $\mu$ being the nucleon chemical potential 
and $\hat N=\int d^3x\, \bar\psi(x) \gamma^0\psi(x)$ the 
nucleon number operator;
\item setting ${\mathcal N}=1$.
\end{enumerate}
By applying these rules, one gets
\begin{equation}
  \label{eq:A002}
  Z(\beta)=\Tr\, e^{-\beta (\hat H-\mu \hat N)}
  =\int D[\bar\psi,\psi,\sigma]e^{\,\,
    \int\limits_0^\beta d\tau \int d^3x\, {\mathcal L}_E(\tau,{\bf x})}~.
\end{equation}
The Euclidean Lagrangian ${\mathcal L}_E$ is obtained by replacing 
$t$ with $-i\tau$ in the Minkowskian Lagrangian \eqref{eq:H001} and reads
\begin{equation}
  \label{eq:D003}
  \begin{split}
    {\mathcal L}_E(\tau,\mathbf x)\equiv {\mathcal L}_E(\bar x)&=
    \int d^4\bar y\, \overline{\psi}(\bar x)S^{-1}_{0E}(\bar x-\bar y)
    \psi(\bar y) \\
    &+\frac{1}{2}\int d^4\bar y\, \sigma(\bar x) 
    D^{-1}_{0E}(\bar x-\bar y)\sigma(\bar y)
    +g\, \overline{\psi}(\bar x)\sigma(\bar x)\psi(\bar x)  
    -U[\sigma(\bar x)] \\
    &\equiv-\left[{\mathcal H}(\bar x)
      -\mu\, \overline{\psi}(\bar x)\gamma^0\psi(\bar x)\right]~, 
  \end{split}
\end{equation}
where we have denoted the Euclidean space--time coordinates with an upper bar
[$\bar x=(\tau,\mathbf x)$, $\bar y=(\tau_y,\mathbf y)$] 
and ${\mathcal H}$ is the (Euclidean) Hamiltonian density.
In the zero temperature limit $S^{-1}_{0E}$ and $D^{-1}_{0E}$
are given by
\begin{eqnarray} 
  S^{-1}_{0E} (\bar x-\bar y) &=& -(\gamma^0\partial_{\tau_x}+i\, 
  {\boldsymbol \gamma}\cdot
  {\mathbf \nabla}_x+m)\, \delta^4(\bar x-\bar y)~, \\
  D^{-1}_{0E} (\bar x-\bar y) &=& 
  -(-\partial^2_{\tau_x}-{\mathbf \nabla}_x^2+m^2_\sigma)\, 
  \delta^4(\bar x-\bar y)~.
\end{eqnarray}
The Euclidean propagators in momentum space follow
from Eqs.~(\ref{eq:019}),(\ref{eq:A008}) with the replacements
$k_0\to ik_0+\mu$ and $q_0\to iq_0$ and take the form
\begin{eqnarray}
\label{eq:007}
  S_{0E}(k)&=&\frac{1}{\gamma_0(ik_0+\mu)-{\boldsymbol \gamma}\cdot{\bf k}-m}~, \\
\label{eq:A028}
  D_{0E}(q)&=&-\frac{1}{q_0^2+{\bf q}^2+m_\sigma^2}~.
\end{eqnarray}

Once the partition function $Z(\beta)$ is known, the ground 
state ($T=0$) energy of 
nuclear matter follows from the relation
\begin{equation}
  \label{eq:014}
  E\equiv \langle \hat H \rangle
  =\lim_{\beta\to \infty}\left(-\frac{\partial}{\partial \beta} \log Z(\beta)
  +\mu A\right)=
 \lim_{\beta\to \infty}\left(-\frac{\log Z(\beta)}{\beta}+ \mu A\right)~,
\end{equation}
$A\equiv \langle \hat N \rangle$ being the total number of nucleons.
Note that the nucleon chemical potential is related to the Fermi momentum by
\begin{equation}
  \label{eq:D002}
  \lim_{\beta\to \infty} \mu=E_F^*\equiv\sqrt{k_F^2+(m^*)^2}~,
\end{equation}
where $m^*$ is the nucleon effective mass at the mean field level 
[see Eq.~\eqref{effm}] and follows from 
the dynamics underlying the Lagrangian \eqref{eq:H001}.
In the next Section we shall introduce an approximate scheme to evaluate 
the ground state energy and therefore the binding energy per nucleon:
\begin{equation}
  \label{eq:010}
  \frac{B.E.}{A}\equiv \frac{E}{A}-m\equiv\frac{\epsilon}{\rho}-m~,
\end{equation}
where $\epsilon\equiv E/\Omega$ and $\rho\equiv A/\Omega=2k_F^3/(3\pi^2)$
are the energy density and the number density of nuclear matter, respectively.
Note that we are implicitly considering 
symmetric nuclear matter ($k_{F_n}=k_{F_p}=k_F$) but all our
results  can be easily adapted to the asymmetric case.

Once  $\epsilon$ is known, other observables are accessible.
Pressure (and hence the equation of state) and compressibility
are indeed given by the thermodynamical relations:
\begin{equation}
  \label{eq:011}
  P=\rho^2\frac{\partial}{\partial \rho}
  \left(\frac{\epsilon}{\rho}\right)=-\epsilon
  +\rho\frac{\partial \epsilon}{\partial \rho}~,
\end{equation}
and
\begin{equation}
\label{eq:011b}
   {\mathcal K}=9\frac{\partial P}{\partial\rho}
   =9\rho\frac{\partial^2\epsilon}{\partial\rho^2}~.
\end{equation}
Other quantities, such as for instance the symmetry energy,
could also be addressed, but they involve the dynamics of 
isovector mesons, which has been recently studied in 
Refs.~\cite{KuKu-97,Li-al-02}, and will be considered by 
us in future investigations.

Within the present path integral method, the problem of evaluating
the static properties of nuclear matter requires the elaboration
of approximation techniques in order to compute the functional integral which
defines the partition function \eqref{eq:A002}.

\section{The Bosonic Effective Action}
\label{sec:3}

As stated in the Introduction, we shall concentrate here on
the approximation scheme given by the Boson Loop 
Expansion~\cite{AlCeMoSa-87,CeCoSa-97}. As a first step to 
derive it, one has to introduce a bosonic effective action
corresponding to the Lagrangian density (\ref{eq:H001}) but no 
longer displaying fermionic degrees of freedom.
This is the specific purpose of the present Section.

At variance with Ref.~\cite{chin}, we present the theory in a heuristic way, 
disregarding in the present Section the renormalization problem and the 
$\sigma$ self--interaction terms 
---which in turn add non-trivial complications to the renormalization problem--- 
and leaving for a second step (Sec.~\ref{sec:4a}) the details of the renormalization 
procedure; the latter, indeed, affects the binding energy in the medium already 
at the mean field level~\cite{chin}. We also notice that the classification of
the counterterms in the usual Loop Expansion and in the BLE is different.

Since we shall focus our attention only on the boson--like observables
of Eqs.~\eqref{eq:014}--\eqref{eq:011b}, we can safely put
$\bar\eta=\eta=0$ in the generating functional of Eq.~\eqref{eq:013}. 
The functional integral is then converted into an integral over bosonic 
variables only, by explicitly integrating out the fermionic field; one obtains
\begin{equation}
  \label{eq:A019}
    Z[J]=\frac{1}{{\mathcal N}}
    \int D[\sigma]\,
    e^{\,iA^B_{\rm eff}[\sigma]+i\int d^4x\, \sigma(x) J(x) } ~.
\end{equation}
By using well known properties of the following Gaussian integral 
over Grassmann variables:
$$\int D[\bar x, x]\;e^{-\bar x A x} =\det A=e^{\Tr \log A}~,$$
the bosonic effective action $A^B_{\rm eff}$ entering into Eq.~\eqref{eq:A019} 
turns out to be
\begin{eqnarray}
  \label{eq:017}
    A^B_{\rm eff}[\sigma]&=&\frac{1}{2}\int d^4x\, d^4y\, \sigma(x) 
    D_0^{-1}(x-y)\sigma(y)-i\Tr\log[-i(S_0^{-1}+g\sigma)]\\
    &=&\frac{1}{2}\int d^4x\, d^4y\, \sigma(x) D_0^{-1}(x-y)\sigma(y)+
    A^B_0+i{\rm Tr}\sum_{n=1}^{\infty}\frac{(-1)^n}{
      n}\left[S_0g\sigma\right]^n \nonumber \\
    &=&\frac{1}{2}\int d^4x\, d^4y\, \sigma(x) D_0^{-1}(x-y)\sigma(y)+
    A^B_0-i{\rm Tr}\int\limits_0^g\frac{d\lambda}{\lambda}
    \frac{\lambda\sigma}{S_0^{-1}+\lambda\sigma}~. \nonumber
\end{eqnarray}
The term
\begin{equation}
  \label{eq:A009}
  A^B_0=-i\Tr\log[-iS_0^{-1}]
\end{equation}
is related to the functional generator of non--interacting fermions by 
\begin{equation}
  \label{D001}
  e^{iA^B_0}=\int D[\bar\psi,\psi]e^{i\int d^4x\, d^4y\,
  \bar\psi(x) S_0^{-1}(x-y)\psi(y)}~.
\end{equation}
Let us stress that, although the fermionic degrees of 
freedom have been integrated out,
their influence on the generating functional \eqref{eq:A019}
remains unaltered.


Note that in \eqref{eq:017}
space--time integrals are also contained in the traces, in addition 
to a sum over spin and isospin. The trace of the logarithm acquires a meaning 
only through its series expansion (second line of Eq.~\eqref{eq:017}).
Moreover $\Tr[S_0g\sigma]^n$ is a short notation for
\begin{multline}
  \label{eq:020}
  \Tr[S_0g\sigma]^n=4 g^n
  \int d^4x_1\dots d^4x_n\,{\mathfrak P}^{(n)}_0(x_1,\dots,x_n)\sigma(x_1)
  \sigma(x_2)
  \dots \sigma(x_n)~,
\end{multline}
where we have defined
\begin{equation}
  \label{eq:H758}
  {\mathfrak P}^{(n)}_0(x_1,\dots,x_n)=S_0(x_1-x_2)  S_0(x_2-x_3)
  \dots  S_0(x_n-x_1)\,,
\end{equation}
and the factor 4 originates from the spin--isospin trace. 

It will be also useful to introduce the vertices ${\mathfrak P}^{(n)}_0$
evaluated in the vacuum, denoted by ${\mathfrak P}^{(n)}_{0v}$: they are
obtained from Eq.~\eqref{eq:H758} with the replacement $S_0\to S_{0v}$,
while the corresponding density--dependent contributions are
${\mathfrak P}^{(n)}_{0M}= {\mathfrak P}^{(n)}_{0}- {\mathfrak P}^{(n)}_{0v}$.

Their Fourier transforms are also needed. Here 
the independent variables are $n-1$ and we get
\begin{eqnarray}
  \label{eq:H912}
  &&\widetilde{\mathfrak P}^{(n)}_0(p_1,p_2,\dots p_{n-1})= \\
 &&\quad -i\intq{p}S_0(p)S_0(p+p_1)S_0(p+p_1+p_2)\dots
S_0(p+p_1+\dots +p_{n-1})\,.
\nonumber
\end{eqnarray}

The Euclidean version of the bosonic effective action is
\begin{align}
  \label{eq:A0102}
  A^B_{E\,\rm eff}(\beta)&=\Tr\log[-S^{-1}_{0E}]+
  \tilde A^B_{E\,\rm eff}(\beta)~,\\
  \tilde A^B_{E\,\rm eff}(\beta)&=\frac{1}{2}
  \int d^4\bar x\, d^4\bar y\, \sigma(\bar x) D^{-1}_{0E}(\bar x -\bar y)
  \sigma(\bar y)
  -{\rm Tr}\sum_{n=1}^{\infty}\frac{(-1)^n}{
    n}\left[S_{0E}g \sigma\right]^n \nonumber \\
  &=\frac{1}{2}
  \int d^4\bar x\, d^4\bar y\, \sigma(\bar x) D^{-1}_{0E}
  (\bar x -\bar y)\sigma(\bar y)
  +{\rm Tr}\int\limits_0^g\frac{d\lambda}{\lambda}
  \frac{\lambda\sigma}{S_{0E}^{-1}+\lambda\sigma}~, \nonumber
\end{align}
and the partition function takes the form
\begin{equation}
  \label{eq:A010}
  Z(\beta)=
  Z_0(\beta)\int D[\sigma]\, e^{\tilde A^B_{E\,\rm eff}(\beta)}~.
\end{equation}
Here,
\begin{equation}
  \label{eq:D001}
  Z_0(\beta)=\det [-S^{-1}_{0E}]=
  \int D[\bar\psi,\psi]\,e^{\,\,\int d^4\bar x\,d^4\bar y\, 
  \bar\psi(\bar x) S_{0E}^{-1}(\bar x - \bar y)\psi(\bar y)}
\end{equation}
is the partition function for an assembly of non--interacting 
nucleons and the remaining functional integral in Eq.~\eqref{eq:A010} 
represents a system of self--interacting bosons.

The interaction terms of Eq.~\eqref{eq:020} 
are built up with closed fermionic loops, either 
in the Minkowskian or Euclidean space, and play
a central role in the present treatment.
It has been shown in Refs.~\cite{CeSa-88,CeCoCoSa-92}
that, within a non--relativistic
kinematics, they can be either evaluated explicitly or at least reduced to the 
evaluation of a one--dimensional integral. Note however 
that the difficulties met in Refs.~\cite{CeSa-88,CeCoCoSa-92} 
mainly derived from the kinematical singularities
in Eq.~\eqref{eq:020}. The use of the Euclidean metric shifts all of them
in the complex plane, considerably simplifying the numerical integrations.
Since these interaction terms
represent the building blocks of the BLE, they require a careful 
treatment. Relativistic calculations are much more sensitive to the 
dynamics of the mesons with respect to non--relativistic case. 
In the latter, indeed, at least in most practical cases, the mesonic dynamics
is frozen (the meson exchange is reduced to a static potential) and it
is reflected in some spin operators which are easily traced out.
On the contrary, in a relativistic approach
convective currents  cannot be neglected  without violating 
Lorentz covariance. We shall discuss 
in the next Section how some archetypal cases may be constructed and how
one can handle more complicated dynamics.

To evaluate the generating functional \eqref{eq:A019} or  the partition
function \eqref{eq:A010}, we adopt the semiclassical expansion.
Its lowest order, i.e., the mean field level,
corresponds to the stationary phase approximation 
(saddle point approximation in the Euclidean space). At this stage, one 
requires the bosonic action to be stationary with respect to small 
variations
of the field $\sigma$:
\begin{equation}
\label{stationary}
\frac{\delta A^B_{\rm eff}[\sigma]}{\delta \sigma}=0~.
\end{equation}
The following equation of motion  is thus 
obtained:
\begin{equation}
  \label{eq:021}
  \int d^4y\, D_0^{-1}(x-y)\sigma(y)=
  -i\Tr\sum_{n=1}^\infty (-1)^n[S_0g \sigma]^{n-1}S_0 g
  =i g \, \Tr\, \frac{S_0}{1+S_0g\sigma}~.
\end{equation}
One recognizes that $\sigma$ cannot be vanishing at the mean 
field level, since the first term of the sum in Eq.~\eqref{eq:021} 
is just a fermionic line closed onto itself, namely the tadpole
$$ig\Tr\, S_0=4ig\int d^4y\, S_0(y-y)~.$$
The above quantity is obviously divergent but the usual renormalization 
techniques, described in Sec.~\ref{sec:2emezzo}, make it
 finite and non--trivial~\cite{chin,AlCeMoSa-88}.
 This occurrence  emphasizes the peculiar role 
played by the scalar--isoscalar field, which is indeed the only one
 generating a tadpole, even in the vacuum, where
any other meson field has vanishing expectation value.

For static, infinite nuclear matter, the mean field solution
is uniform in space and time, thus the lhs of the field 
equation~\eqref{eq:021} reduces to
\begin{equation}
\int d^4y\, D^{-1}_0(x-y)\sigma(y)=
-\left(\Box_x+m^2_\sigma\right)\sigma(x)=-m^2_\sigma \bar \sigma~.
\end{equation}
In the ``no--sea'' approximation,
the mean field $\bar \sigma$ is then obtained by solving the 
following self--consistency equation (we introduce the scalar density 
$\rho_{s}=\langle \bar \psi \psi\rangle$ and its density--dependent part 
$\rho_{sM}=\langle \bar \psi \psi\rangle_M$):
\begin{equation}
  \label{eq:026}
  \begin{split}
    \bar\sigma&=\frac{g}{m^2_\sigma} \rho_{sM}
       =-i\frac{g}{m_\sigma^2}\Tr\, S_{H M}=
     -4i\frac{g}{m_\sigma^2}\intq{k}S_{H M}(k)\\
    &=\frac{g}{m_\sigma^2}\frac{m^*}{\pi^2}\left(k_F E_F^*
      -(m^*)^2\log\frac{k_F+E_F^*}{m^*}\right)~,
\end{split}
\end{equation}
where $S_{H M}$ is the density--dependent part of
the Hartree--dressed nucleon propagator
\begin{equation}
  \label{eq:023}
  S_H=\frac{1}{S_0^{-1}+g\bar\sigma}~,
\end{equation}
that is obtained from the definition of $S_0$ in Eq.~\eqref{eq:019} after 
replacing $m$ with the nucleon effective mass:
\begin{equation}
\label{effm}
m^*=m-g\bar\sigma \,.
\end{equation}

\begin{figure}[ht]
  \begin{center}
    \epsfig{file=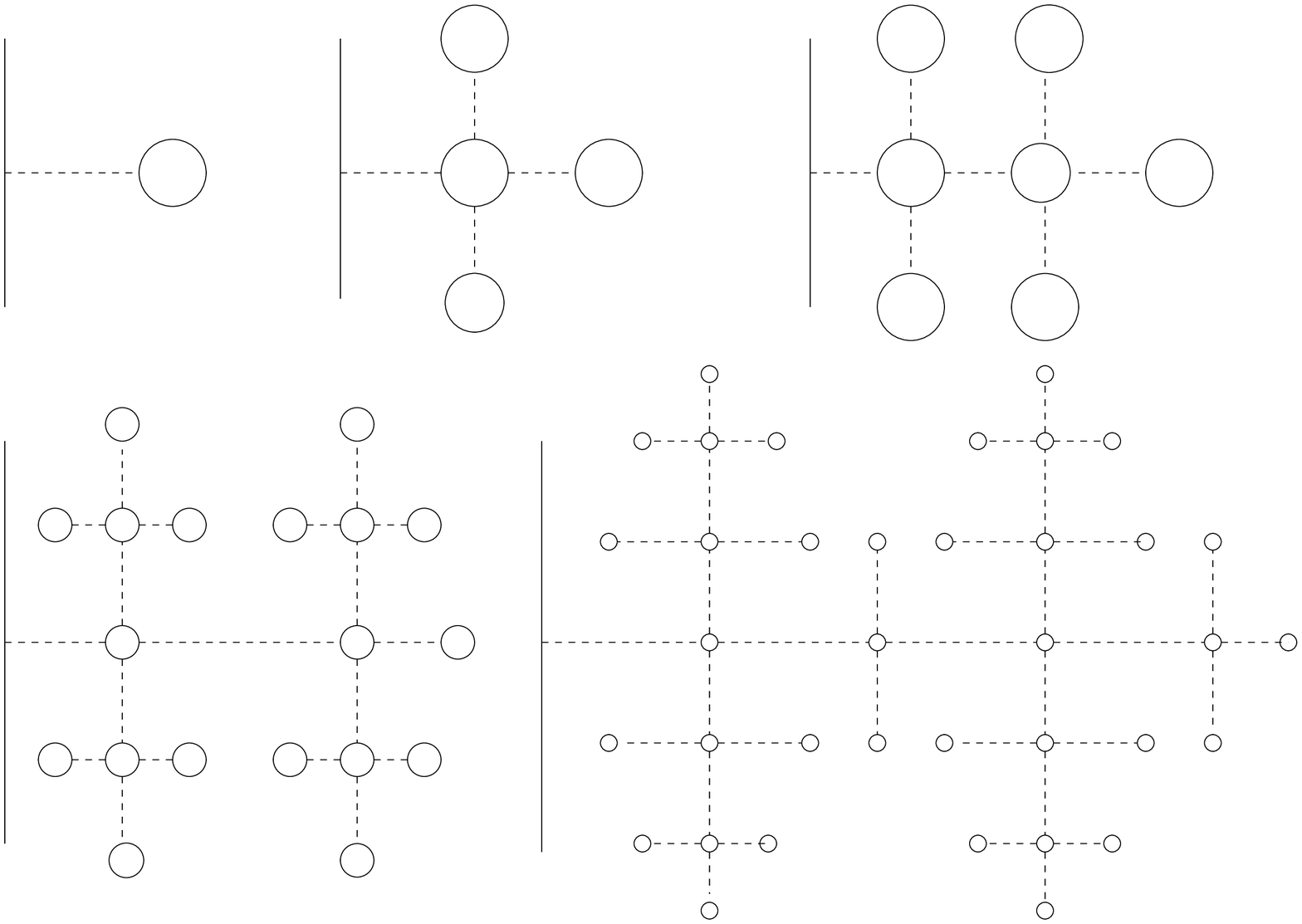,width=10cm,height=7cm}
    \caption{Mean field level nucleon self--energy  
      diagrams for $\sigma$ exchange.}
    \label{fig:2}
  \end{center}
\end{figure}

From a diagrammatic point of view, self--consistency
amounts to account for diagrams like those of Figure~\ref{fig:2}, in which 
each nucleon propagator is dressed by the Hartree self--energy.

In view of an application of the present scheme to QHD, one has to consider
the contribution of the vector--isoscalar meson $\omega$ as well.
The $\omega$ mean field is obtained as 
$\bar\omega_\mu=({g_\omega}/{m_\omega^2})\rho\, \delta_{\mu0}~,$
without any self--consistency condition because the nuclear density 
$$\rho\equiv \langle \bar\psi \gamma_0 \psi \rangle=2\frac{k^3_F}{3\pi^2}$$
is now implied instead of the scalar density $\rho_s$.
The relativistic (Lorentz contraction) effect stemming from
the difference between the two mean values
was already noted long time ago by Lee and Wick \cite{LeWi-74} and 
ultimately entails a restoration of the chiral symmetry at sufficiently 
high  density in the Wigner mode.

Since Eq.~\eqref{eq:026} provides a non--vanishing solution for 
the mean field $\bar \sigma$, 
in order to apply the standard technique of the Semiclassical approximation
one needs to define a new scalar--isoscalar field
$\sigma'(x)=\sigma(x)-\bar\sigma$ with vanishing expectation value 
at the mean field level, $\bar{\sigma'}=0$.
By rewriting the generating functional \eqref{eq:A019} with 
the above change of variable and then renaming $\sigma'$ as $\sigma$,
the bosonic effective action \eqref{eq:017} becomes
\begin{equation}
  \label{eq:030}
   A^B_{\rm eff}[\sigma]=\frac{1}{2}\sigma D_0^{-1}\sigma-\frac{m_\sigma^2}{2}
   (2\sigma\bar\sigma+{\bar\sigma}^2)+A^B_0
   +i{\rm Tr}\sum_{n=1}^\infty\frac{(-1)^n}{n}[S_0g(\sigma+\bar\sigma)]^n
\end{equation}
and the coupling to the external field changes to $(\sigma+\bar\sigma)J$.
\footnote{Note that a compact notation has been used for the first two terms
of the effective action \eqref{eq:030}. Their explicit expression reads:
\[ \frac{1}{2}\int d^4x\, d^4y\, \left[\sigma(x)D^{-1}_0(x-y)\sigma(y)
-m^2_\sigma \delta^4(x-y)\left(2\sigma(x) \bar \sigma + {\bar \sigma}^2
\right)\right]~.\] }

Next, one can write
\begin{multline}
  \label{eq::530}
  i{\rm Tr}\sum_{n=1}^\infty\frac{(-1)^n}{n}[S_0g(\sigma+\bar\sigma)]^n\\
  =i{\rm Tr}\sum_{n=1}^\infty\frac{(-1)^n}{n}[S_Hg\sigma]^n
  +i{\rm Tr}\sum_{n=1}^\infty\frac{(-1)^n}{n}[S_0g\bar\sigma]^n~,
\end{multline}
as it can be easily verified keeping in mind Eq.~\eqref{eq:023} and 
by rewriting the series in terms of the
corresponding $\log$'s. By using Eq.~\eqref{eq:026}, the $n=1$ term in the
first sum of the rhs of Eq.~\eqref{eq::530} reads 
$-ig\Tr S_H\sigma = m_\sigma^2 \sigma \bar\sigma~,$ 
so that no linear term in $\sigma$ enters the effective action:
\begin{multline}
  \label{eq:630}
   A^B_{\rm eff}[\sigma]=\frac{1}{2}\sigma D_0^{-1}
   \sigma-\frac{m_\sigma^2}{2}\bar\sigma^2\\
   +A^B_0+i{\rm Tr}\sum_{n=2}^\infty\frac{(-1)^n}{n}[S_H g \sigma]^n
   -i\Tr\log[1+S_0 g \bar\sigma]~.
\end{multline}
Let us remark that the constant terms in $A^B_{\rm eff}$ are
irrelevant in evaluating the generating functional,
but they are significant in  the partition function. By using
the definition \eqref{eq:A009}, one can recognize the following reordering
\begin{equation}
\label{const}
A^B_0-i\Tr\log[1+S_0 g \bar\sigma]=-i\Tr\log[-iS_H^{-1}]~.
\end{equation}

We then consider the partition function. The part of the Euclidean action
originating from the constant term \eqref{const} now
replaces the factor $Z_0(\beta)$ of Eq.~\eqref{eq:D001}. 
The difference between Eq.~\eqref{const}
and Eq.~\eqref{eq:A009} lies in the replacement $m\to m^*$,
namely $S_{0E}\to S_{HE}=(S^{-1}_{0E}+g\bar \sigma)^{-1}$, 
$S_{HE}$ being the Euclidean version of $S_{H}$.
Therefore we can write 
\begin{equation}
  \label{eq:A015}
  Z(\beta)=Z_0^*(\beta)\int D[\sigma]\, 
  e^{\,\widehat A^B_{E\,\rm eff}(\beta)}~,
\end{equation}
with 
\begin{equation}
  Z^*_0(\beta)=\det [-S^{-1}_{HE}]=
\int D[\bar\psi,\psi]\, e^{\int d^4\bar x\,d^4\bar y\, \bar\psi(\bar x) 
S_{HE}^{-1}(\bar x - \bar y)\psi(\bar y)}~,
\end{equation}
and
\begin{equation}
  \label{eq:A022}
  \widehat A^B_{E\,\rm eff}(\beta)=\frac{1}{2}\sigma D^{-1}_{0E}\sigma
  -\Omega\beta\frac{m_\sigma^2}{2}\bar\sigma^2
  -{\rm Tr}\sum_{n=2}^{\infty}\frac{(-1)^n}{
    n}\left[S_{HE}\, g \sigma\right]^n~.
\end{equation}
In this expression, space--time integrations are understood for the 
free $\sigma$ contribution, but in the second, constant term in the rhs, the 
time ($\beta$) and volume ($\Omega$) integrations have been made explicit.

\section{The elementary vertices of the bosonic action}
\label{sec:4x}

We analyze now the elementary vertices corresponding to $\Tr[S_Hg\sigma]^n$
and entering the bosonic effective action \eqref{eq:630}. 
When embedded in a Feynman diagram, a term $\Tr[S_Hg\sigma]^n$ 
is represented by a fermionic loop
with $n$ external points; a boson propagator can be attached 
to each one of these points, including the spin and isospin matrices
which pertain to the specific boson. 
Isospin traces only factorize out a real number, but the possible presence
of $\gamma$ matrices, as it is the case for the $\omega$ exchange,
entails a non--trivial dependence upon the momenta, which is usually
neglected in non--relativistic calculations. In this section
we only consider some archetypal structures. More realistic
situations need to be dealt with individually.

Let us first fix the kinematics. A loop with $N+1$ incoming or outgoing
mesons will depend upon $N$ momenta $p_k$, while $p_{N+1}=-\sum_{k=1}^{N}p_k$
is fixed by momentum conservation. To simplify the notation, it is convenient 
to define $N$ auxiliary momenta according to
\begin{eqnarray}
  q_0&=&0~, \\
  q_i&=&\sum_{j=1}^{i}\, p_j \;\;\;\; (i=1,\dots, N)~. \nonumber
\label{qi}
\end{eqnarray}

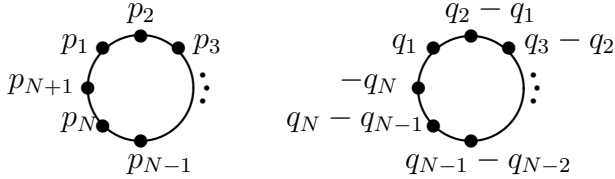
\begin{figure}[h]
\begin{center}
\begin{minipage}[c]{325pt}
\message{Figure 1}
\begin{picture}(320,100)(20,0)
\thicklines
\put(110,50){\circle{40}}
\put(95.7,65.3){\circle*{5}}
\put(80,65.3){$p_1$}
\put(110,70){\circle*{5}}
\put(105,77){$p_2$}
\put(124.3,65.3){\circle*{5}}
\put(130,65.3){$p_3$}
\put(133.5,55){\circle*{1.5}}
\put(135,50){\circle*{1.5}}
\put(133.5,45){\circle*{1.5}}
\put(110,30){\circle*{5}}
\put(105,20){$p_{N-1}$}
\put(95.7,35.7){\circle*{5}}
\put(80,35.7){$p_N$}
\put(90,50){\circle*{5}}
\put(60,50){$p_{N+1}$}

\put(235,50){\circle{40}}
\put(220.7,65.3){\circle*{5}}
\put(205,65.3){$q_1$}
\put(235,70){\circle*{5}}
\put(225,77){$q_2-q_1$}
\put(249.3,65.3){\circle*{5}}
\put(255,65.3){$q_3-q_2$}
\put(258.5,55){\circle*{1.5}}
\put(260,50){\circle*{1.5}}
\put(258.5,45){\circle*{1.5}}
\put(235,30){\circle*{5}}
\put(210,20){$q_{N-1}-q_{N-2}$}
\put(220.7,35.7){\circle*{5}}
\put(165,35.7){$q_{N}-q_{N-1}$}
\put(215,50){\circle*{5}}
\put(185,50){$-q_{N}$}
\end{picture}
\protect\caption{{The kinematics of the generic interaction term in the expansion
of the bosonic effective action \eqref{eq:630}.}
\label{fig:A001}}
\end{minipage}
\end{center}
\end{figure}

The kinematics is clarified in Figure~\ref{fig:A001}. 
With these definitions the generic vertex in momentum space reads:
$$\intq{p}{\mathcal O}_0(p,q_0)S_H(p+q_0)
{\mathcal O}_1(p,q_1)S_H(p+q_1)\dots{\mathcal O}_N(p,q_N)S_H(p+q_N)~,$$
where the ${\mathcal O}_i$ are matrices whose particular form depends 
on the bosonic field(s) one considers.

Actually, even the simplest case, namely the exchange of scalar--isoscalar 
mesons, is rather involved, because the traces over the $\gamma$ matrices are
non--trivial. 

In order to handle this case, we introduce a set of auxiliary functions as follows.
First we rewrite the nucleon propagator in nuclear matter with Fermi momentum 
$k_F$ as
$$S_{0}(k)=(\rlap/k+m) {\mathfrak S}_0(k)~,$$
where
\begin{equation}
  \label{eq:E001}
  {\mathfrak S}_0(k)=\frac{1}{2E_k}\left\{\frac{\theta(|{\bf k}|-k_F)}{k_0-E_k+i\eta}+
    \frac{\theta(k_F-|{\bf k}|)}{k_0-E_k-i\eta}
      -\frac{1}{k_0+E_k-i\eta}\right\}~.
\end{equation}
Then we introduce the auxiliary functions
\begin{equation}
  \label{eq:E002}
  \Pi^{(N+1)}(q_0,\dots,q_N)=-i\intq{p}\prod_{i=0}^N{\mathfrak S}_0(p+q_i)~,
\end{equation}
which, at variance with the ${\mathfrak P}^{(n)}_0$ of Eq.~(\ref{eq:H758}), 
do not contain the matrix structure of the propagators $S_0$.

Following Refs.~\cite{CeSa-88,CeCoCoSa-92} we integrate over $p_0$
by closing the integration path with a half circle in the lower half plane, 
getting
\begin{eqnarray}
  \label{eq:E003}
    &&\Pi^{(N+1)}(q_0,\dots,q_N)=-\sum_{j=0}^N\intt{p}
    \prod_{i=0 \atop i\not=j}^N
    \frac{\theta(|{\bf p}+{\bf q}_j|-k_F)}{4E_{p+q_j}E_{p+q_i}}\\
    && \hskip 3.8cm \times \left\{\frac{\theta(|{\bf p}+{\bf q}_i|-k_F)}
      {q_{i0}-q_{j0}-(E_{p+q_i}-E_{p+q_j})}\right. \nonumber \\
    &&\left.+\frac{\theta(k_F-|{\bf p}+{\bf q}_i|)}
      {q_{i0}-q_{j0}-(E_{p+q_i}-E_{p+q_j})-i\eta}
    -\frac{1}{q_{i0}-q_{j0}+E_{p+q_i}+E_{p+q_j}-i\eta}\right\}~. \nonumber
\end{eqnarray}
The rhs displays a sum of products. Each term in the product
is in turn a sum of three pieces, with poles at
$q_{i0}-q_{j0}-(E_{p+q_i}-E_{p+q_j})=0$ (first term), 
$q_{i0}-q_{j0}-(E_{p+q_i}-E_{p+q_j})-i\eta=0$ (second term) and
$q_{i0}-q_{j0}+E_{p+q_i}+E_{p+q_j}-i\eta=0$ (third term).
In the first term, at variance with the others,  the singularity is removable
and hence no prescription on how to handle it is needed. Indeed this singularity 
arises if we consider the $i$--th term of the product in the $j$--th term of the sum; 
but, by exchanging $i$ and $j$, a term with the same singularity and opposite sign 
appears, thus removing the singularity. Being irrelevant, we can formally add 
to the denominator of the first term an infinitesimal factor 
$+i\eta\, {\rm sign}(q_{i0}-q_{j0})$. In the second term the $\theta$ functions 
impose $E_{p+q_i}-E_{p+q_j}<0$ and the denominator can vanish only when 
$q_{i0}-q_{j0}<0$. This allows to replace the $-i\eta$ by 
$+i\eta\, {\rm sign}(q_{i0}-q_{j0})$ and to combine together  
the first two terms, the $\theta$ functions summing up to 1. 
Finally, the same factor can be ascribed \emph{a fortiori} to the imaginary part
of the last term, thus yielding the compact expression:
\begin{eqnarray}
  \label{eq:E005}
    \Pi^{(N+1)}(q_0,\dots,q_N)&=&-\sum_{j=0}^N
    \intt{p}\prod_{i=0\atop i\not=j}^N
    \frac{\theta(|{\bf p}+{\bf q}_j|-k_F)}{2E_{p+q_j}}\\
    &&\times \frac{1}{\left[q_{i0}-q_{j0}+E_{p+q_j}+i\eta\, 
        {\rm sign}(q_{i0}-q_{j0})
        \right]^2-E_{p+q_i}^2}~. \nonumber
\end{eqnarray}
If we are only interested to the in--medium contribution of the above vertices 
we have to subtract from \eqref{eq:E005} its value in the vacuum; this leads to the
following expression, where no residual divergence is left:
\begin{eqnarray}
  \label{eq:E006}
    \Pi^{(N+1)}(q_0,\dots,q_N)&=&\sum_{j=0}^N\intt{p}\prod_{i=0\atop i\not=j}^N
    \frac{\theta(k_F-|{\bf p}+{\bf q}_j|)}{2E_{p+q_j}}\\
    &&\times\frac{1}{\left[q_{i0}-q_{j0}+E_{p+q_j}
        +i\eta\, {\rm sign}(q_{i0}-q_{j0})
        \right]^2-E_{p+q_i}^2}~. \nonumber
\end{eqnarray}
The transition to the Euclidean space is even simpler, because in 
the replacement $q_{i0}\to iq_{i0} +\mu$ the chemical potential cancels out, 
while the $+i\eta\,{\rm sign}(q_{i0}-q_{j0})$ becomes irrelevant. One gets:
\begin{eqnarray}
  \label{eq:E007}
    \Pi_E^{(N+1)}(q_0,\dots,q_N)&=&\sum_{j=0}^N\intt{p}
    \prod_{i=0\atop i\not=j}^N
    \frac{\theta(k_F-|{\bf p}+{\bf q}_j|)}{2E_{p+q_j}}\\
    &&\times\frac{1}{\left[iq_{i0}-iq_{j0}+E_{p+q_j}
        \right]^2-E_{p+q_i}^2}~. \nonumber
\end{eqnarray}

Relativity forces in the elementary vertices of the bosonic effective action 
many complications with respect to the non--relativistic case discussed in
Refs.~\cite{CeSa-88,CeCoCoSa-92}, and no general 
formula can be given in the relativistic case. 
In Ref.~\cite{BaCeQu-05} a detailed study of the two--point vertices
has been carried out. Here we shall illustrate with a simple, but yet 
non--trivial, example how the three--point vertex can be reduced to the
evaluation of auxiliary functions $\Pi^{(N+1)}(q_0,\dots,q_N)$. 
Let us consider the emission of three $\sigma$ mesons from a 
three--point fermionic loop. The corresponding vertex  is
\begin{eqnarray}
  \label{eq:E016}
  {\mathfrak P}^{(3)}(q_1,q_2)&\equiv&-i\intq{p}\\
   &&\times \frac{\Tr(\rlap/p+m)
    (\rlap/p+\rlap/q_1+m)(\rlap/p+\rlap/q_2+m)}{[p^2-m^2\pm i\eta]
    [(p+q_1)^2-m^2\pm i\eta][(p+q_2)^2-m^2\pm i\eta]} \nonumber
\end{eqnarray}
(the sign of the $i\eta$ follows from the above discussion).
By performing the traces, the numerator of the integrand becomes
$$4m^3+12mp^2+8mp\cdot q_1+8mp\cdot q_2+4mq_1\cdot q_2~,$$
and with some algebra can be recast in the form
\begin{multline*}
4m\bigl\{(4m^2-q_1^2-q_2^2+q_1\cdot q_2)\\+(p^2-m^2)+
[(p+q_1)^2-m^2]+[(p+q_2)^2-m^2]\bigr\}~.
\end{multline*}
By inserting this result in Eq.~\eqref{eq:E016} we then get
\begin{eqnarray}
  \label{eq:E017}
  {\mathfrak P}^{(3)}(q_1,q_2)&=&4m(4m^2-q_1^2-q_2^2+q_1\cdot q_2)
  \Pi^{(3)}(q_1,q_2)\\
  &&+4m\Pi^{(2)}(q_2-q_1)+4m\Pi^{(2)}(q_1)+4m\Pi^{(2)}(q_2)~. 
  \nonumber
\end{eqnarray}
Hence in this particular case a realistic three--point vertex is reduced to 
the evaluation of the auxiliary vertices $\Pi^{(3)}$ and $\Pi^{(2)}$.
The procedure can be systematically generalized.

\section{Renormalization of the Bosonic Effective Action}
\label{sec:2emezzo}

The derivation of Section \ref{sec:3}  only  has a formal 
meaning since no renormalization procedure has been applied up to now. 
To make the above derived quantities meaningful we need \\
(1) to add to the Lagrangian a set of counterterms and \\
(2) to fix, by altering the Feynman rules, a regularization procedure.

In the present case it is convenient to rewrite the Lagrangian \eqref{eq:H001}
for bare quantities, denoted by the index $0$: 
\begin{equation}
  \label{eq:H031}
  {\mathcal L}=\overline{\psi}_0(i\rlap/\partial-m_0)\psi_0
  +\frac{1}{2}\partial_\mu\sigma_0\, \partial^\mu \sigma_0
  -\frac{1}{2}m_{\sigma 0}^2\sigma_0^2+g_0\overline{\psi}_0\sigma_0\psi_0
  -\frac{\lambda_{30}}{3!}\sigma_0^3-\frac{\lambda_{40}}{4!}\sigma^4_0~. 
\end{equation}
Bare fields and constants are  linked 
to the renormalized ones by the relations:
\begin{subeqnarray}
  \label{eq:H032}
  \psi_0&=&Z_2^\frac{1}{2}\psi\\
  \sigma_0&=&Z_3^\frac{1}{2}\sigma\\
  m_0&=&m+Z_2^{-1}\delta m\\
  m_{\sigma 0}^2&=&m_\sigma^2+Z_3^{-1}\delta m_{\sigma}^2\\
  g_0&=&Z_1Z_2^{-1}Z_3^{-\frac{1}{2}}g\\
  \lambda_{30}&=&Z_4Z_3^{-\frac{3}{2}}\lambda_3\\
  \lambda_{40}&=&Z_5Z_3^{-2}\lambda_4~.
\end{subeqnarray}
By substituting we rewrite the Lagrangian in the form
\begin{equation}
  \label{eq:H033}
{\mathcal L}=  
{\mathcal L}_{\rm ren}+{\mathcal L}_{ct}
 ={\mathcal L}_0+{\mathcal L}_I+{\mathcal L}_{\rm ct}~,
\end{equation}
where
\begin{subeqnarray}
  \label{eq:A034}
  {\mathcal L}_0&=&\overline{\psi}(i\rlap/\partial-m)\psi
  +\frac{1}{2}\partial_\mu\sigma\, \partial^\mu \sigma
  -\frac{1}{2}m_\sigma^2\sigma^2\\
  {\mathcal L}_I&=&g\overline{\psi}\sigma\psi
  -\frac{\lambda_3}{3!}\sigma^3-\frac{\lambda_4}{4!}\sigma^4\\
  {\mathcal L}_{\rm ct}&=&-\delta m\bar\psi\psi
  +(Z_2-1)\bar\psi(i\rlap/\partial-m)\psi\\
  &&-\frac{1}{2}\delta m_\sigma^2\sigma^2+(Z_3-1)
  \left\{\frac{1}{2}\partial_\mu\sigma\, \partial^\mu \sigma
    -\frac{1}{2}m_\sigma^2\sigma^2\right\}\nonumber\\
  &&+(Z_1-1)g\bar\psi\sigma\psi-(Z_4-1)\frac{\lambda_3}{3!}\sigma^3
  -(Z_5-1)\frac{\lambda_4}{4!}\sigma^4~. \nonumber
\end{subeqnarray}

According to the general theory of renormalization \cite{It-Zu-80,Collins}, each order
of the perturbative expansion remains now finite. It is convenient for our 
purposes to fix only \emph{one} expansion parameter, the coupling $g$; 
this can be achieved by putting
$\lambda_3=m_\sigma \alpha_3g^3$ and $\lambda_4=\alpha_4g^4$ and
attributing to $\delta m$, $\delta m_\sigma^2$, $(Z_2-1)$ and $(Z_3-1)$ the order
$g^2$, to $(Z_1-1)$ and $(Z_4-1)$ the order $g^3$ and to $(Z_5-1)$
the order $g^4$. 

Let us now fix a regularization procedure: its
choice in many--body calculations is largely immaterial, but for the present
purposes and for practical calculations it will be useful 
to fix a scheme. We shall simply adopt a cutoff scheme, according to
\begin{equation}
  \label{eq:H270}
  \intq{p}\,\dots\longrightarrow \intq{p}\,\dots\,
  \left(\frac{\Lambda^2}{p^2-\Lambda^2+i\eta}\right)^2
\end{equation}
(the square is just the minimal required power to force 
convergence in any of the diagrams occurring in the theory).

Pauli--Villars regularization \cite{It-Zu-80} 
will also be kept in mind, since it will help
us in the power counting inside the fermion loops.
The dimensional regularization adopted in Ref.~\cite{chin}, instead, 
appears less natural in the present many--body context; indeed the nuclear medium
breaks Lorentz covariance and it becomes unclear to which  dimension 
(3--dim space or 1--dim time) the regularization has to be applied.

A short discussion of a few possible renormalization schemes, 
although well known \cite{It-Zu-80}, is required in this context. 
\begin{enumerate}
\item One can choose the parameters in Eqs.~(\ref{eq:A034}a)--(\ref{eq:A034}g)
  as the {\em physical} masses and coupling constants,
  so that at each perturbative order a counterterm (infinite or finite) must
  be introduced to keep them fixed (hard renormalization). This entails
  that the renormalization point must be on--mass--shell for any external
  line in evaluating the relevant diagrams. 
\item One can instead  remove from any elementarily
  divergent diagram its divergent part (soft renormalization).
  Dimensional regularization is usually adopted within this scheme.
\item Intermediate choices are also possible. For instance Ref.~\cite{chin}
  chooses the renormalization point at $q^2=0$ (thus generating some 
  inconsistency in the definition of the $\sigma$ mass).
\item Finally, as suggested in Ref.~\cite{AlCeMoSa-88}, one can adopt as counterterms 
  the values of the elementarily divergent diagrams in the vacuum, considered 
  as functions of the incoming and outgoing momenta. A rigorous treatment of the
  renormalization is then lost, since the ill--definition of the Green's functions 
  is no longer restricted to a point, but the resulting scheme is finite. Moreover
  one disentangles, at least partially,  the physics of the nucleon, lying in the
  form of the vertex functions in the vacuum, from the nuclear physics. 
  When this scheme, referred to 
  as ``vacuum subtraction'' scheme in the following, is applied to QHD, one of the
  main advantages is to avoid meaningless contributions from the vacuum. One should 
  indeed keep in mind that QHD is intrinsically an effective theory, with 
  parameters ruled by nuclear dynamics. 
\end{enumerate}

A detailed discussion of the elementarily divergent diagrams is now in order.
In fact the procedure does not follow the usual path of perturbation theory, 
since loops containing or not a boson line play different roles and 
the requirement of being one--particle irreducible (1PI) looses its validity in the
context of the BLE (in fact it refers now only to boson lines). 
Furthermore the tadpoles (referred to in the following
as Hartree contributions) play a special role.

\begin{figure}[th]
  \begin{center}
    \leavevmode
    \epsfig{file=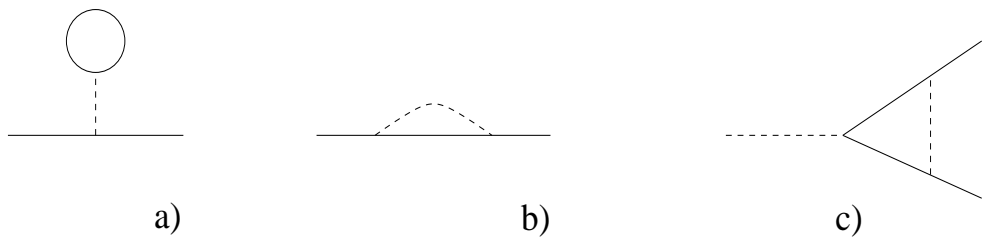,height=3cm,width=13cm}
    \caption{a) The Hartree contribution to the nucleon
      mass renormalization. b) and c) The elementarily divergent diagrams 
with one internal boson line.}
    \label{fig:ct_1}
  \end{center}
\end{figure}

We consider first the Hartree contribution to the nucleon self--energy of 
Fig.~\ref{fig:ct_1}. 
It plays a peculiar role in the BLE, since it will 
eventually disappear after the self--consistency requirement is forced; thus it does
not increase the bosonic loop number, neither the fermionic one. It reads
\begin{equation}
  \label{eq:H503}
  \Sigma_{Hv}=-\frac{g^2}{m_\sigma^2}(-i)\Tr\intq{q}S_{0v}(q)=-
  \frac{g^2}{m_\sigma^2}\rho_{s\, v}(m)~,
\end{equation}
where $\rho_{s\, v}=\langle \bar \psi \psi\rangle|_{k_F=0}$
is the scalar density in the vacuum. We have emphasized its dependence upon
$m$ since, as already heuristically anticipated in Section \ref{sec:3}, 
the replacement of $m$ with the effective mass is expected.
The explicit expression of $\rho_{s\, v}(m)$ within the regularization scheme
given by Eq.~\eqref{eq:H270} is provided in the Appendix.

Next we consider the divergent diagrams with one bosonic loop. They are
displayed in Fig.~\ref{fig:ct_1} and represent the Fock term of the 
self--energy (the name being borrowed from the standard many--body
language) and the $\sigma NN$ vertex correction.
The diagrams a) and b) of Fig.~\ref{fig:ct_1}
remove the divergences from the nucleon self--energy.

In full generality, in the vacuum 
the inverse propagator of the nucleon can be written in the form
(we use the standard notation for the 1PI
vertices and the corresponding generator, $\Gamma$)\footnote{We 
indicate the two-- and more--points vertex functions in momentum space with
a short--hand notation, which implies a Fourier transform. For example:
\[ \frac{\delta^2 \Gamma}{\delta \bar \psi \delta \psi} \equiv
\frac{1}{(2\pi)^4}\int d^4(x_1-x_2) e^{ip\cdot(x_1-x_2)}
\frac{\delta^2 \Gamma}{\delta \bar \psi(x_1)\delta \psi(x_2)} \]
}:
\begin{equation}
  \label{eq:H501}
    \Gamma_{\bar\psi\psi}(p)\equiv
\frac{\delta^2 \Gamma}{\delta \bar \psi \delta \psi}
=Z_2(\rlap/p-m)-\delta m-\Sigma_v(p)~,
\end{equation}
where the self--energy in the vacuum takes the form:
\begin{equation}
  \label{eq:H811}
  \Sigma_v(p)=\Sigma^{(0)}+(\rlap/p-m)\Sigma^{(1)}+(\rlap/p-m)^2
  \Sigma_{\rm c}(p)~,
\end{equation}
$\Sigma^{(0)}$ and $\Sigma^{(1)}$
being constant divergent quantities while $\Sigma_{\rm c}(p)$ is 
a convergent function.
Thus hard renormalization entails:
\begin{subeqnarray}
  \label{eq:H901}
  \delta m&=&-\Sigma^{(0)}~,\\
  Z_2-1&=&\Sigma^{(1)}~.
\end{subeqnarray}

Our scheme naturally separates the Hartree term
(already discussed) and the Fock one: we get for the self--energy
\begin{equation}
  \label{eq:H502}
  \Sigma_v(p)=\Sigma_{Hv}+\Sigma_{Fv}(p)+{\rm convergent~ terms}~.
\end{equation}

Similarly we rewrite the mass counterterm as
\begin{equation}
  \label{eq:H115}
  \delta m=\delta m_H+\delta m_F+{\mathcal O}(\hbar^2)~,
\end{equation}
where the ${\mathcal O}(\hbar^2)$ terms are finite and
\begin{equation}
  \label{eq:H701}
  \delta m_H=-\Sigma_{Hv}~.
\end{equation}
Since $\Sigma_{Hv}$ is independent of the momentum,  it
does not contribute to $Z_2-1$, at variance with $\Sigma_{Fv}$.
The latter can be evaluated analytically but its complicated
expression is not given since it is not used in the following.

Next we come to the diagram c) of Fig.~\ref{fig:ct_1}. Again, an explicit 
evaluation of this diagram is not required. The 
corresponding vertex function in the vacuum has the structure
\begin{equation}
  \label{eq:H961}
  \Gamma_{\bar\psi\psi\sigma}(p,q)\equiv
\frac{\delta^3 \Gamma}{\delta \bar \psi \delta \psi \delta \sigma}
    =\widetilde\Gamma^{(c)}+\Gamma^{(c)}_{\rm c}(p,q)\,,
\end{equation}
where $\widetilde\Gamma^{(c)}$ is a logarithmically divergent constant
and $\Gamma^{(c)}_{\rm c}(p,q)$ is a finite function, vanishing at the 
renormalization point. Its choice is however a remarkable source of
ambiguity, as we shall see for the $\sigma^3$ vertex. In the present case 
 we can safely choose to put all particles on--mass--shell.
Furthermore, the explicit form of $\Gamma^{(c)}_{\rm c}(p,q)$ 
is irrelevant in the ``vacuum subtraction'' scheme, since this amounts to neglect 
it at all.
With the above definitions we get
\begin{equation}
  \label{eq:H973}
  Z_1-1=\widetilde\Gamma^{(c)}~.
\end{equation}

The previous considerations complete the renormalization program for the nucleonic 
part of the action. Now we consider the integration over the fermionic fields in the 
generating functional:
\begin{equation}
  \label{eq:H251}
  Z[J]=\frac{1}{\mathcal N}\int D[\bar \psi, \psi, \sigma]\;e^{i\int d^4x 
    [{\mathcal L}_{\rm ren}(x)+ {\mathcal L}_{\rm ct}(x)+ \sigma(x)J(x)]}~.
\end{equation}
This requires special care because it involves divergent quantities, which in addition 
are of different order in the BLE: we must keep $\delta m_H$ in the 0$^{\rm th}$ 
order, being the only fermionic counterterm without mesonic loops, and leave the 
remaining counterterms inside the fermion loops that constitute the effective vertices 
of the bosonic action.

 To remove the explicit occurrence of $\delta m_H$ in the fermionic loops, 
the usual trick is to shift the scalar field as follows
\begin{equation}
  \label{eq:H252}
  \sigma\rightarrow\hat\sigma\equiv\sigma+\frac{\delta m_H}{g}
  =\sigma+\frac{g}{m_\sigma^2}\rho_{s\, v}~
\end{equation}
and replace in the total Lagrangian 
${\mathcal L}_{\rm ren}+ {\mathcal L}_{\rm ct}$ $\sigma$ with $\hat\sigma$:
\begin{multline}
  \label{eq:H253}
  {\mathcal L}=Z_2\bar\psi(i\rlap/\partial  -m)\psi
  -\delta m_F\bar\psi\psi+(Z_1-1)\delta m_H\bar\psi\psi
  +Z_1g\bar\psi\sigma\psi\\
  +\frac{Z_3}{2}
  \left(\partial_\mu\sigma\partial^\mu\sigma-m_{\sigma}^2\hat\sigma^2\right)
  -\frac{1}{2}\delta m_\sigma^2 \hat\sigma^2
  -\frac{1}{3!}Z_4\lambda_3\hat\sigma^3-\frac{1}{4!}Z_5\lambda_4\hat\sigma^4~.
\end{multline}
We can now integrate over the fermionic fields to get the bosonic 
effective action
\begin{multline}
  \label{eq:H254}
  A^{B\, \rm ren}_{\rm eff}[\sigma]=\int d^4x\;\left\{\frac{Z_3}{2}
    \left(\partial_\mu{\sigma}\partial^\mu{\sigma}-m_{\sigma}^2\hat\sigma^2\right)
    -\frac{1}{2}\delta m_\sigma^2 \hat\sigma^2
    -\frac{1}{3!}Z_4\lambda_3\hat\sigma^3-\frac{1}{4!}Z_5\lambda_4\hat\sigma^4\right\}\\
  -i\Tr\log\left\{-i\left[Z_2S_0^{-1}-\delta m_F+(Z_1-1)\delta m_H
  +Z_1g\sigma\right]\right\}~.
\end{multline}

As already shown in Section~\ref{sec:3}, the log 
in the previous action contains the closed fermionic loops, representing the 
effective vertices of our theory. They still contain, however, divergent counterterms, 
which need to be eliminated, as it is illustrated below.

Along the same path of Section~\ref{sec:3} we rewrite [using the symbol $A_0^B$ 
of Eq.~(\ref{eq:A009})]
\begin{multline}
  \label{eq:H260}
  -i\Tr\log\left\{-i\right[Z_2S_0^{-1}-\delta m_F+(Z_1-1)\delta m_H+Z_1g\sigma
    \left]\right\}\\
 = A_0^B -i\Tr\log\left\{1+S_0\left[-\delta m_F+(Z_2-1)S_0^{-1}+(Z_1-1)\delta m_H
  +Z_1g\sigma\right]\right\} \\
  =A_0^B  +i\sum_{n=1}^{\infty}\frac{(-1)^n}{n}
  \Tr\left\{S_0\left[-\delta m_F+(Z_2-1)S_0^{-1}+(Z_1-1)\delta m_H
  +Z_1g\sigma\right]\right\}^n~. 
\end{multline}

\begin{figure}[h]
  \begin{center}
    \leavevmode
    \epsfig{file=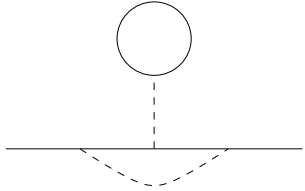,height=2.5cm,width=4cm}
    \caption{The tadpole with vertex correction.}
    \label{fig:tad}
  \end{center}
\end{figure}

Let us notice that $-\delta m_F+(Z_2-1)S_0^{-1}$ is the counterterm canceling 
the divergent part of the diagram b) in Fig.~\ref{fig:ct_1}.
Moreover the existence of the tadpole carries another contribution, 
$(Z_1-1)\,\delta m_H$, which compensates the divergence of the diagram 
of Fig.~\ref{fig:tad}.

The last line of Eq.~\eqref{eq:H260} requires some caution, since the appearance of
the conterterms  $\delta m_F$, $Z_2-1$  and $(Z_1-1)\delta m_H$ seems to alter the
order $n$ of fermionic propagators $S_0$ or, equivalently, of bosonic vertices. 
To clarify this point let us consider the Pauli--Villars regularization \cite{It-Zu-80}, 
which amounts to subtract from a diagram 
containing a $\sigma$ exchange (e.g. diagram b) of Fig.~\ref{fig:ct_1})
a similar process with a heavy meson with mass $\Lambda$. The number of bosonic 
vertices remains obviously unaltered until one takes the limit $\Lambda\to\infty$, 
which shrinks the insertion to one point only, corresponding to a counterterm. 
On this basis, as far as the number of bosonic vertices is concerned, one has to 
count  twice the counterterm insertions
of the type $\delta m_F$ and $Z_2-1$ and, on the same foot, thrice the 
counterterms of the type $Z_1-1$, because diagram c) of Fig.~\ref{fig:ct_1} 
has 3 external points. This amounts to say that $\delta m_F$ and $Z_2-1$ 
are of order $g^2$  and $Z_1-1$ is of order $g^3$.

\begin{figure}[h]
  \begin{center}
    \leavevmode
    \epsfig{file=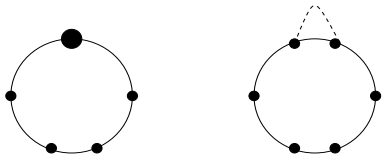,height=2cm,width=5cm}
    \caption{Renormalization of the diagram b) insertions.}
    \label{fig:vert}
  \end{center}
\end{figure}

The above considerations enable us to correctly count the number of 
bosonic vertices (namely of powers of $g$) in a fermionic loop like the one 
illustrated in the right of Fig.~\ref{fig:vert}: 
indeed a loop with, say, $n-2$ $\sigma$--vertices and a counterterm
of the kind $-\delta m_F+(Z_2-1)S_0^{-1}$ (left diagram in Fig.~\ref{fig:vert})
is again of order $g^n$. 
Within the ``vacuum subtraction" scheme, the counterterms are the elementarily 
divergent diagrams evaluated in the vacuum; then the sum of the two diagrams of 
Fig.~\ref{fig:vert} reduces to the diagram on the right with the self--energy
insertion regularized. The same occurs for
the vertex corrections of diagram c) of Fig.~\ref{fig:ct_1} and 
of Fig.~\ref{fig:tad}.

The above procedure proves the renormalizability of the fermionic loops 
appearing in the last line of Eq.~(\ref{eq:H260}) only for $n\geq 5$.
Indeed for lower $n$'s the fermionic loop \emph{itself} is divergent and must 
be handled separately, by subtracting from these fermion
loops their value  at $k_F=0$.

Let us then introduce the following notations:
\begin{equation}
  \label{eq:H978}
  {\mathfrak T}^{(n)}_{0\, M (v)}=g^n(-i)\, {\rm tr}
  \int dx_1\dots dx_n\;{\mathfrak P}^{(n)}_{0\, M (v)}(x_1,\dots,x_n)
  \sigma(x_1)\dots\sigma(x_n)~,
\end{equation}
\begin{eqnarray}
  \label{eq:H744}
  A^B_{M}&=&\sum_{n=1}^\infty\frac{(-1)^n}{n}
  {\mathfrak T}^{(n)}_{0M}~,\\
  A^B_{v}&=&\sum_{n=5}^\infty\frac{(-1)^n}{n}
   {\mathfrak T}^{(n)}_{0v}~,
\end{eqnarray}
where the trace (the symbol {\rm tr}) only refers to spin--isospin indeces.
With the above definitions the bosonic action reads
\begin{multline}
  \label{eq:H979}
    A^{B\,\rm ren}_{\rm eff}[\sigma]=\int d^4x\;\left\{\frac{Z_3}{2}
    \left(\partial_\mu\sigma\partial^\mu\sigma-m_{\sigma}^2\hat\sigma^2\right)
    -\frac{1}{2}\delta m_\sigma^2 \hat\sigma^2 \right. \\
    \left. -\frac{1}{3!}Z_4\lambda_3\hat\sigma^3
    -\frac{1}{4!}Z_5\lambda_4\hat\sigma^4\right\}
  +\sum_{n=1}^4\frac{(-1)^n}{n}
   {\mathfrak T}^{(n)}_{0v}+A^B_{v}+A^B_{M}~,
\end{multline}
where the last two terms are now free from infinities.

\begin{figure}[h]
  \begin{center}
    \leavevmode
    \epsfig{file=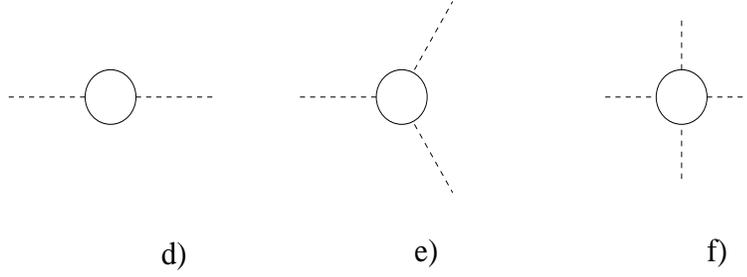,height=4cm,width=10cm}
    \caption{The elementarily divergent diagrams contributing to the 
renormalization of d) the $\sigma$ meson mass and wave function and 
e) the $3\sigma$ and f) $4\sigma$ vertices.}
    \label{fig:ct_3}
  \end{center}
\end{figure}

We then proceed to fix the renormalization parameters 
$\delta m_\sigma^2$, $Z_3$, $Z_4$ and $Z_5$ in order to 
remove the remaining infinities. This goal is achieved in perturbation theory,
starting from the meson self--energy diagram d) of Fig.~\ref{fig:ct_3},
which renormalizes the $\sigma$ mass and coupling constant.
The corresponding two--point vertex reads:
\begin{equation}
  \label{eq:H5011}
  \Gamma_{\sigma\sigma}(q)\equiv \frac{\delta^2 \Gamma}{\delta \sigma^2}
  =\frac{Z_3}{2}(q^2-m_\sigma^2)-
  \frac{1}{2}\delta m_\sigma^2    
  -\frac{1}{2}\Pi_v^{(d)}(q^2)~,
\end{equation}
where
\begin{equation}
  \label{eq:H905}
  \begin{split}
    \Pi^{(d)}_{v}(q^2)&=ig^2\Tr\widetilde{\mathfrak P}_{0\,v}^{(2)}(q)\\
    &=\Pi^{(d)}_{v}(m_\sigma^2)+(q^2-m_\sigma^2)\frac{d\Pi^{(d)}_{v}(q^2)}{dq^2}
    \Bigm|_{q^2=m_\sigma^2}+(q^2-m_\sigma^2)^2\Pi^{(d)}_{c}(q^2)~.
  \end{split}
\end{equation}
$\Pi_c^{(d)}(q^2)$ being a convergent function. Renormalization then requires:
\begin{subeqnarray}
  \label{eq:H903}
  \delta m_\sigma^2  &=&-\Pi_v(m_\sigma^2)\\
  Z_3-1&=&\frac{d\Pi_v(q^2)}{dq^2}
  \Bigm|_{q^2=m_\sigma^2}~.
\end{subeqnarray}

Let's consider in details the divergent contribution of diagram d) 
of Fig.~\ref{fig:ct_3}.
By  applying Eqs.~\eqref{eq:H903} we see that in Eq.~(\ref{eq:H979}) 
$\int d^4x [\frac{1}{2}(Z_3-1)\left(\partial_\mu\sigma\partial^\mu\sigma
-m_{\sigma}^2\sigma^2\right)-\frac{1}{2}\delta m_\sigma^2 \sigma^2]$ and 
$(1/2) {\mathfrak T}^{(2)}_{0v}$ cancel each other up to finite terms
which are vanishing on--shell as well as in the ``vacuum subtraction'' scheme.

Explicit formulas for $\Pi^{(d)}(m_\sigma^2)$ and 
$\frac{d\Pi^{(d)}(q^2)}{dq^2}\Bigm|_{q^2=m_\sigma^2}$ are given in the Appendix
together with their expressions both in the hard and
in the soft renormalization scheme, the latter being adopted in Ref.~\cite{chin},
with renormalization point at $q^2=0$. As we shall see, the
difference is relevant.

In Fig.~\ref{fig:ct_3} the elementarily divergent diagrams
contributing to the 3$\sigma$ [diagram e)] and 4$\sigma$ [diagram f)] 
vertices are also depicted. We define:
\begin{eqnarray}
  \label{eq:H910}
  \Gamma_{\sigma\sigma\sigma}(p,q)&\equiv& 
  \frac{\delta^3 \Gamma}{\delta \sigma^3}
  =\widetilde\Gamma^{(e)}+
  \Gamma^{(e)}_{\rm c}(p,q)~,\\
  \label{eq:H9100}
  \Gamma_{\sigma\sigma\sigma\sigma}(p,q,k)&\equiv&
  \frac{\delta^4 \Gamma}{\delta \sigma^4}
  =\widetilde\Gamma^{(f)}+
    \Gamma^{(f)}_{\rm c}(p,q,k)~,
\end{eqnarray}
where the quantities $\widetilde\Gamma^{(e)}$ and $\widetilde\Gamma^{(f)}$
are logarithmically divergent (in $\Lambda$) constants and $\Gamma^{(e)}_{\rm c}$ 
and $\Gamma^{(f)}_{\rm c}$ are suitable finite functions
which disappear in  the ``vacuum subtraction'' scheme.
Renormalization enforces
\begin{eqnarray}
  \label{eq:H991}
  \lambda_3(Z_4-1)&=&\widetilde\Gamma^{(e)}~,\\
  \lambda_4(Z_5-1)&=&\widetilde\Gamma^{(f)}~,
\end{eqnarray}
and the divergent diagram e) [f)] of Fig.~\ref{fig:ct_3}
exactly cancels the ${\mathfrak T}^{(3)}_{0v}$
$[{\mathfrak T}^{(4)}_{0v}]$ contribution in the action (\ref{eq:H979}).

\begin{figure}[h]
  \begin{center}
    \leavevmode
    \epsfig{file=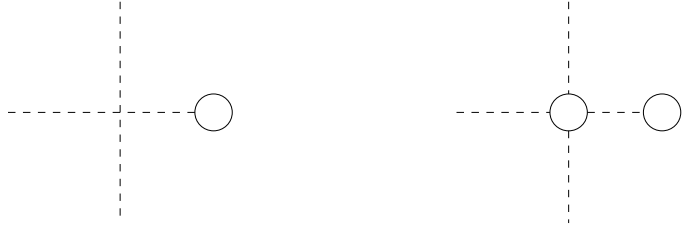,height=3cm,width=9cm}
    \caption{Further divergent diagrams contributing to 
      $\Gamma_{\sigma\sigma\sigma}$.}
    \label{fig:ct_5}
  \end{center}
\end{figure}

\begin{figure}[h]
  \begin{center}
    \leavevmode
    \epsfig{file=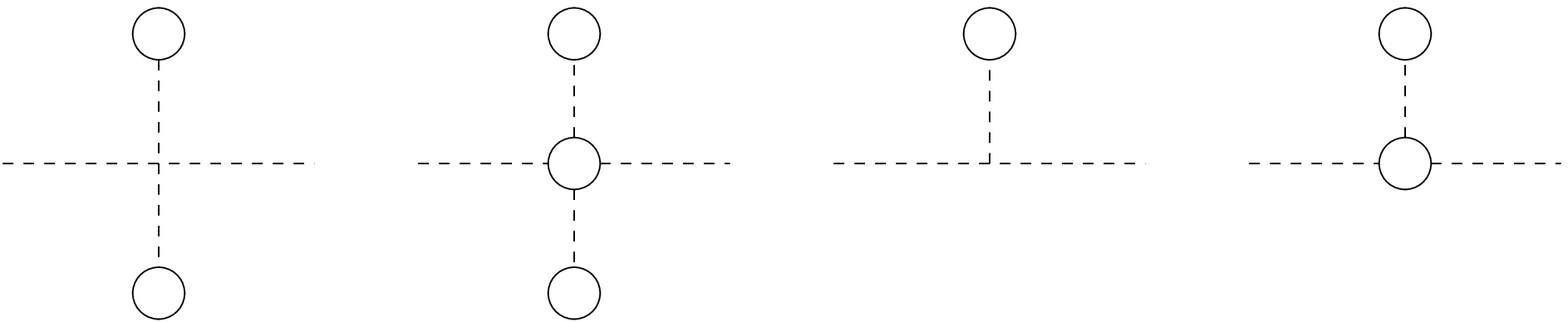,height=2.5cm,width=12.5cm}
    \caption{Further divergent diagrams contributing to 
      $\Gamma_{\sigma\sigma}$.}
    \label{fig:ct_6}
  \end{center}
\end{figure}

\begin{figure}[h]
  \begin{center}
    \leavevmode
    \epsfig{file=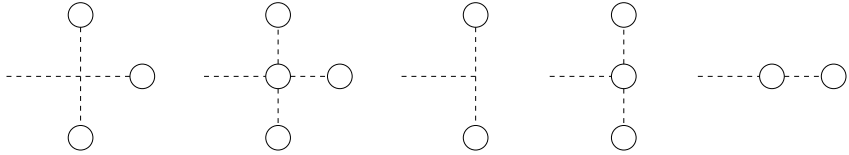,height=2cm,width=11.2cm}
    \caption{Divergent diagrams contributing to the linear part in $\sigma$
      of the bosonic effective action.}
    \label{fig:ct_7}
  \end{center}
\end{figure}

The above discussion does not exhaust the cancellation of infinities present
in the action \eqref{eq:H979}; other, non 1PI, divergent diagrams exist.
This is linked to a subtlety in proving, via a L\'egendre transformation
\cite{Amit}, that the functional $\Gamma$ is the generator of the
1PI diagrams. A careful examination of this proof reveals
that it does not apply to the tadpoles like $\Sigma_{Hv}$. 
For physical processes in the vacuum
this does not create any pathology, since
$\Sigma_{Hv}$ is automatically renormalized to 0 for $k_F=0$
and no further divergences survive in \eqref{eq:H979}.
However, in the present context, the tadpole in the medium is neither
vanishing nor trivial, and some care is required. Indeed we must
also account for those diagrams that are {\em not} 1PI due to
the presence of one or more $\Sigma_{Hv}$ insertions. They contribute
not only to $\Gamma_{\sigma\sigma}$ and $\Gamma_{\sigma\sigma\sigma}$
but also to the term of the action which is linear in $\sigma$ and
are shown in Figs.~\ref{fig:ct_5}--\ref{fig:ct_7}.
 We note that these diagrams are indeed originated by the quadratic, cubic
and quartic terms in the shifted scalar field
$\hat\sigma=\sigma- \Sigma_{Hv}/g$ appearing in the effective action \eqref{eq:H979}.
When these diagrams are accounted for, taking care of their multiplicity,
one sees that all the remaining divergences in the action cancel out.

Let us consider in  detail the terms of the action linear in $\sigma$. 
The counterterms corresponding to the divergent diagrams of
Fig.~\ref{fig:ct_7} cancel all these terms but one, namely 
$-m_\sigma^2\sigma\left(\frac{\delta m_H}{g}\right)$. The latter, however,
is canceled by the term ${\mathfrak T}^{(1)}_{0v}$.
Thus, the mean field solution in the vacuum is $\sigma(x)=0$, as it should.

Yet one consideration before leaving the problematic in the vacuum concerns 
the existence of many more divergent diagrams, which contain closed bosonic lines. 
However, the vacuum subtraction technique clearly renormalizes these contributions 
to zero.
Finally, the renormalized bosonic effective action reads
\begin{multline}
  \label{eq:H993}
    A^{B\, \rm ren}_{\rm eff}[\sigma]=\int d^4x\;\left\{\frac{1}{2}
    \left(\partial_\mu\sigma\partial^\mu\sigma-m_{\sigma}^2\sigma^2\right)
    -\frac{1}{3!}\lambda_3\sigma^3-\frac{1}{4!}\lambda_4\sigma^4\right\}\\
  +A^B_{v}+A^B_{M}+{\rm finite~terms}~.
\end{multline}
The only difference with respect to Eq.~\eqref{eq:017}
(besides the presence of the $\sigma$ self--couplings) is that
$\sum_{n=1}^4\frac{(-1)^n}{n} [S_0 g \sigma]^n$
is now replaced by constant, finite terms whose structure depends 
on the choice of the renormalization point. 

Before proceeding to study
the self--consistency equation for the scalar field
we observe that it is possible, as it was done in Ref.~\cite{chin}, 
to put these finite terms to 0 by simply choosing the
renormalization point at $q^2=0$. However, in so doing the $\sigma$ mass 
has to be re--evaluated order by order in the renormalization process.
This would seriously plague a realistic theory (like, say,
a model for the $N N\pi$ interaction) where the masses of the mesons 
as well as the scattering lengths are measured.
In the context of QHD the mass of the $\sigma$ is not measured but plays
the role of a phenomenological parameter which is fixed to
reproduce the saturation properties of nuclear matter. Hence it is irrelevant whether 
the renormalization procedure shifts it, or not.
 There is however an important caveat: one could wonder if 
it has to be fixed before or after the renormalization process.
We shall see that this point is of special relevance.

 We now derive the equation of motion for the mean field.
Looking as usual to a constant, uniform solution for this equation we get
\begin{equation}
  \label{eq:L001}
  \begin{split}
    &\frac{\delta A^{B\,\rm ren}_{\rm eff}[\sigma]}{\delta\sigma}=0\qquad
    \Longleftrightarrow\\
    &-m_\sigma^2\sigma-\frac{\lambda_3}{2}\sigma^2-\frac{\lambda_4}{6}\sigma^3
    -i g \Tr\intq{q}\frac{1}{S_0^{-1}(q)+g\sigma}\\
    &+i\sum_{n=1}^4 (-1)^n g^n \Tr\intq{q}
    S_{0v}^n(q)\sigma^{n-1}=0~,
  \end{split}
\end{equation}
where the last two terms in the lhs of the second equation are 
both divergent but their divergencies cancel each other.
Within our cutoff regularization scheme:
\begin{equation}
  \label{L002}
  -i\Tr\intq{q}\frac{1}{S_0^{-1}(q)+g\sigma}=\rho_s(m-g\sigma)~,
\end{equation}
where the explicit expression of the scalar density $\rho_s$ is given in the Appendix.

The remaining finite contribution in Eq.~(\ref{eq:L001}) can be evaluated by
using the results of the Appendix and taking the limit
$\Lambda\to\infty$. By
denoting with $\bar\sigma$ the solution of Eq.~\eqref{eq:L001} and
recalling the effective mass definition \eqref{effm} we get the 
self--consistency equation
\begin{equation}
  \label{eq:L003}
  -m_\sigma^2\bar\sigma-\frac{\lambda_3}{2}\bar\sigma^2-\frac{\lambda_4}{6}\bar\sigma^3
  +\Sigma_{\rm sc}(\bar\sigma)=0~,
\end{equation}
where
\begin{eqnarray}
\label{sigma-chin}
  &&\Sigma_{\rm sc}(\bar\sigma)\equiv g \rho_s(m-g\bar\sigma)
  +i\sum_{n=1}^4 (-1)^n g^n \Tr\intq{q}
    S_{0v}^n(q)\bar\sigma^{n-1}= \\
    &&-\frac{g}{\pi^2}\left\{(m-g\bar\sigma)^3
      \log\frac{k_F+E^*_F}{m}-(m-g\bar\sigma)k_FE_F^*-m^2\bar\sigma
      -\frac{5}{2}m\bar\sigma^2-\frac{11}{6}\bar\sigma^3\right\} \nonumber
\end{eqnarray}
and $E_F^*=\sqrt{k_F^2+(m-g\bar\sigma)^2}$.

 Before concluding this discussion we wish to stress once more
that the choice of the renormalization point introduces some ambiguity. Indeed 
it is an aspect of a more general question, namely the relevance that
the quantum vacuum effects should, or should not, have on the model.

In the previous derivation
we have split the interaction brought in the bosonic action
by the fermion loops into two parts: the density--dependent and vacuum
contributions. We could reasonably ask whether the effect of the vacuum 
(the presence of the Dirac sea) should pertain to the particle or to the
nuclear world. This matter is something more than a merely philosophical
question, since it affects the self--consistency equation in
a crucial way. If one indeed adopts the vacuum subtraction scheme, one simply gets
\begin{equation}
  \label{eq:L004}
  \Sigma_{\rm sc}(\bar\sigma)= g \rho_{sM}(m-g\bar\sigma)~.
\end{equation}
This eliminates the corrections (Relativistic Hartree Approximation)
introduced in Eq.~(\ref{sigma-chin}) and formally derived for the first time 
by Chin \cite{chin}. At the same time, the suggestion that the treatment of the
vacuum dynamics in the approach of Ref.~\cite{chin} is unsatisfactorily 
handled~\cite{Fur-97}  seems to find a natural explanation.

On the other hand, if one accepts the soft renormalization scheme of Ref.~\cite{chin},
at each perturbative order the
$\sigma$ mass varies   and its true value 
 is reached when a fixed point, ruled by the renormalization group 
equation, is met. We can conclude that in  the vacuum subtraction scheme, 
the $\sigma$ mass is fixed by hard renormalization; in this case the 
  Relativistic Hartree contribution introduced by Chin must
  be neglected since it is an effect due to the
  dynamics in the vacuum.

\section{The Bosonic Loop Expansion}
\label{sec:4a}
 
In order to explicitly evaluate the partition function \eqref{eq:A015}, 
which through Eqs.~\eqref{eq:014}--\eqref{eq:011b} supplies the static 
properties of nuclear matter, an approximation scheme is required.
As in Refs.~\cite{AlCeMoSa-87,CeCoSa-97}, we use a Semiclassical expansion 
at the leading and next--to--leading orders: this amounts to evaluate 
first the partition function at the saddle point and then the
quadratic quantum mechanical fluctuations around the saddle 
point. The $\sigma$ self--interaction terms are
disregarded in the formal derivations of the present Section.

The standard Semiclassical approximation consists in an expansion in
powers of $\hbar$. By making explicit the dependence on this
constant, one realizes that a factor $\hbar^{-1}$ 
appears in front of the original action for fermions and
bosons [see Eq.~\eqref{eq:013}]. However, after integration over the fermionic
fields, Planck's constant also appears with a non-trivial dependence inside the new bosonic 
effective action of Eqs.~\eqref{eq:630} and \eqref{eq:A022}. 
Thus, the use of the stationary phase
approximation does not lead to the standard 
Semiclassical expansion in powers of $\hbar$.
In this case, the ordinary way to proceed \cite{Amit,NeOr-88} is to introduce a
dimensionless parameter $a^{-1}$ ($a$ being small)
in front of the bosonic effective action, to expand in powers of $a$
and finally to set 
$a=1$ at the end of the calculation at any given order in $a$.

\begin{figure}[ht]
  \leavevmode
  \begin{center} 
    \epsfig{file=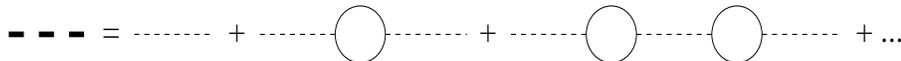,width=12cm,height=0.8cm}
    \caption{Dynamical content of the RPA--dressed boson propagator
      $\Delta^{\rm RPA}$: dashed lines means 
      free $\sigma$ propagators, solid circles represent fermionic loops.}
    \label{fig:3} 
\end{center} 
\end{figure}

The bosonic effective action of Eq.~\eqref{eq:630} [equivalent to  
the Euclidean action \eqref{eq:A022}] has been constructed in such a way
that the corresponding mean field equation 
only admits the solution $\sigma=0$: no term linear in $\sigma$ is 
present in this action. 
Thus, we first separate in $\widehat A^B_{E\,\rm eff}$
of Eq.~\eqref{eq:A022} the terms quadratic in $\sigma$ from the higher order ones.
By applying this procedure, from Eq.~\eqref{eq:A015}
we obtain the partition function in the form\footnote{The introduction of the
term with the external source $J$ will be useful for later manipulations
of the functional integral (see Eq.~\eqref{eq:A035}
and Section~\ref{2l}). The need for the factor $\sqrt{a}$ 
in front of the external source will become clear later.}
\begin{equation}
  \label{eq:A024}
  Z(\beta)[J]=Z_0^*(\beta)\, e^{-\Omega\beta\frac{m_\sigma^2}{2}\bar\sigma^2}
  \int D[\sigma]\, e^{\frac{1}{a}\left[
    A^{B0}_{E\,\rm eff}[\sigma]+ A^{BI}_{E\,\rm eff}[\sigma]
    +\sqrt{a}\int d^4\bar x\, \sigma(\bar x) J(\bar x)\right]}~.
\end{equation}
In the above, $A^{B0}_{E\,\rm eff}$ contains the part of the effective
action which is quadratic in $\sigma$ and $A^{BI}_{E\,\rm eff}$ the 
remaining effective interaction:
\begin{equation}
  \label{eq:A027}
  A^{BI}_{E\,\rm eff}[\sigma]=-{\rm Tr}\sum_{n=3}^{\infty}\frac{(-1)^n}{
    n}\left[S_{HE}\, g \sigma\right]^n~.
\end{equation}
More explicitly, a first contribution to $A^{B0}_{E\,\rm eff}$ comes from the free 
bosonic field while a second one originates from the first term of the 
effective interaction in Eq.~\eqref{eq:A022} and reads
\begin{eqnarray}
  \label{eq:A026}
  -\frac{1}{2}\Tr [S_{HE}\, \sigma]^2&=&
  -\frac{1}{2}\int d^4\bar x\,d^4\bar y\,\sigma(\bar x) 
   \, {\rm tr}\left[S_{HE}(\bar x-\bar y)S_{HE}(\bar y-\bar x)\right]\sigma(\bar y) \\
  &\equiv& -\frac{1}{2}\int d^4\bar x\,d^4\bar y\,
  \sigma(\bar x)\Pi_{0E}(\bar x-\bar y)\sigma(\bar y)~, 
\end{eqnarray}
which implicitly defines the equivalent of the Lindhard function
in the relativistic Euclidean case, $\Pi_{0E}$.  
The trace in the rhs of Eq.~\eqref{eq:A026} (the symbol {\rm tr})
only refers to spin--isospin indexes; in this case it
just amounts to an implicit factor 4. Thus:
\begin{equation}
  \label{eq:A128}
  A^{B0}_{E\,\rm eff}[\sigma]=\frac{1}{2}\sigma\left(D_{0E}^{-1}-\Pi_{0E}
    \right)\sigma \equiv \frac{1}{2}\sigma (\Delta^{\rm RPA}_E)^{-1}\sigma~,
\end{equation}
where we have introduced the Euclidean Random Phase Approximation (RPA) 
propagator
\begin{equation}
  \label{eq:A029}
  \Delta^{\rm RPA}_E=\left(D_{0E}^{-1}-\Pi_{0E} \right)^{-1}~,
\end{equation}
with the diagrammatic representation given in Figure~\ref{fig:3}.
Actually, only direct ring contributions enter into this propagator.

\subsection{Mean Field}

The mean field contribution ($\sigma=0$)
to the partition function \eqref{eq:A024} comes from the constant in front of 
the functional integral:
\begin{equation}
  \label{eq:A031}
  Z(\beta)[J]\bigm|_{\rm mean~field}=Z_0^*(\beta)
  e^{-\Omega\beta\frac{m_\sigma^2}{2}\bar\sigma^2}~.
\end{equation}
It is well known that
\begin{eqnarray}
  \label{eq:A032}
  Z_0^*(\beta)&=&\int D[\overline{\psi},\psi]\,e^{\int d^4\bar x\, d^4\bar y\,
    \overline{\psi}(\bar x)S^{-1}_{HE}(\bar x - \bar y)\psi(\bar y)} \\
    &=& \prod_{{\mathbf p},\lambda}\left(1+e^{-\beta(E_p^*-\mu)}
    \right)\left(1+e^{-\beta(E_p^*+\mu)}\right)~, \nonumber
\end{eqnarray}
where $E_p^*=\sqrt{{\bf p}^2+{m^*}^2}$, $m^*=m-g \bar\sigma$ 
and $\lambda$ is a spin--isospin
index, is nothing but the partition function of a relativistic free Fermi gas
up to the replacement of the bare nucleon mass with the effective one. 
The ground state energy of nuclear matter at the mean field level is then
deduced from Eq.~\eqref{eq:014}:
\begin{eqnarray}
  \label{eq:A061}
  E\bigm|_{\rm mean~field}&\equiv&
  -\lim_{\beta\to \infty}\frac{\partial \log Z(\beta)[J]\bigm|_{\rm mean~field}}
  {\partial \beta} +E^*_F A \\
  &=&\sum_{{\mathbf p},\lambda} E^*_p\, \theta(E^*_F-E^*_p)+\Omega\frac{m^2_\sigma}
  {2}\bar\sigma^2~. \nonumber
\end{eqnarray}
The well--known mean field result by Serot and Walecka \cite{SeWa-86} for the 
energy density:
\begin{eqnarray}
\epsilon \bigm|_{\rm mean~field}&\equiv& \frac{E\bigm|_{\rm mean~field}}{\Omega}
=4\int \frac{d^3p}{(2\pi)^3}E^*_p\, \theta(k_F-|{\bf p}|)
+\frac{m^2_\sigma}{2}\bar \sigma^2 \\
&=&\frac{1}{4\pi^2}\left\{k_F E_F^*[(m^*)^2+2k_F^2]-(m^*)^4\log 
\frac{k_F+E_F^*}{m^*} \right\} +\frac{m^2_\sigma}{2}\bar \sigma^2~, \nonumber
\end{eqnarray}
easily follows from Eq.~\eqref{eq:A061} 
with the replacement $\sum_{{\mathbf p}, \lambda}/\Omega \to 
4\int d^3p /(2\pi)^3$, the factor 4 being the
spin--isospin degeneracy.

\subsection{One--loop correction}

In order to evaluate the functional integral in Eq.~\eqref{eq:A024}
we make the change of variable $\sigma\to \sqrt{a}\, \sigma$,
thus obtaining
\begin{eqnarray}
  \label{eq:A035}
    Z(\beta)[J]&=&Z_0^*(\beta)e^{-\Omega\beta\frac{m_\sigma^2}{2}\bar\sigma^2} \\
    &&\times \sqrt{a}\int D[\sigma]\, 
         e^{\left[\frac{1}{2}\sigma(\Delta^{\rm RPA}_E)^{-1}\sigma
         -\sum_{n=3}^\infty \frac{(-1)^n}{n}a^{(\frac{n}{2}-1)}\Tr[S_{HE}\, g \sigma]^n
        + \sigma J\right]}\nonumber \\
    &=&Z_0^*(\beta)e^{-\Omega\beta\frac{m_\sigma^2}{2}\bar\sigma^2} \nonumber \\ 
    &&\times \sqrt{a}\, e^{-\sum_{n=3}^\infty
        \frac{(-1)^n}{n}a^{(\frac{n}{2}-1)}\Tr[S_{HE}\, g\, \frac{\delta~}{\delta J}]^n}
      \int D[\sigma]\, e^{\left[\frac{1}{2}\sigma(\Delta^{\rm RPA}_E)^{-1}\sigma
        +\sigma J\right]}\nonumber \\
    &=&Z_0^*(\beta)e^{-\Omega\beta\frac{m_\sigma^2}{2}\bar\sigma^2}
    \left\{\det[-\Delta^{\rm RPA}_E]\right\}^\frac{1}{2} \nonumber \\
    &&\times \sqrt{a}\, e^{-\sum_{n=3}^\infty
      \frac{(-1)^n}{n}a^{(\frac{n}{2}-1)}\Tr[S_{HE}\, g \frac{\delta~}{\delta J}
      ]^n}e^{-\frac{1}{2}J\Delta^{\rm RPA}_E J}~, \nonumber
\end{eqnarray}
where in the last equality a Gaussian integration has been performed\footnote{Note
that the Euclidean RPA propagator $\Delta_E^{\rm RPA}$ is negative--definite:
$$\Delta_E^{\rm RPA}(q)=-1/[q_0^2+{\bf q}^2+m_\sigma^2+\Pi_{0 E}(q)]~.$$}.

The lowest order quantum mechanical correction beyond the mean field
is thus obtained from the partition function:
\begin{equation}
  \label{eq:A062}
  Z(\beta)[J]\bigm|_{\rm 1-loop}=\left\{\det[-\Delta^{\rm RPA}_E]\right\}^\frac{1}{2}~,
\end{equation}
which, as we shall see, corresponds to the one--boson--loop approximation. 
By using Eq.~\eqref{eq:A029}, the one--loop contribution to the 
ground state energy of nuclear matter is obtained from:
\begin{eqnarray}
  \label{eq:A063}
    -\frac{\log Z(\beta)[J]\bigm|_{\rm 1-loop}}{\beta}
    &=&\frac{1}{2\beta}\Tr\log\left(-D_{0E}^{-1}+\Pi_{0E}\right)\\
    &=&\frac{1}{2\beta}\left\{\Tr\log [-D_{0E}^{-1}]
      -\Tr\sum_{n=1}^\infty\frac{1}{n}(D_{0E}\Pi_{0E})^n\right\}\nonumber \\
    &=&\frac{1}{2\beta}\left\{\Tr\log [-D_{0E}^{-1}]-\Tr\int\limits_0^1
      d\lambda\;\frac{D_{0E}\Pi_{0E}}{1-\lambda D_{0E}\Pi_{0E}}\right\}~. \nonumber 
\end{eqnarray}
The first term in the last line is the vacuum energy of the $\sigma$ boson,
which is washed out by renormalization. 
One also identifies the first term of the sum in the second line as the
Fock contribution, in which the two nucleon propagators are Hartree--dressed
(see Figure~\ref{fig:2}).
The ground state energy at the one--loop level turns out to be
\begin{eqnarray}
  \label{eq:A064}
  E\bigm|_{1-\rm loop}&\equiv &
      -\lim_{\beta\to\infty}\frac{\log Z(\beta)[J]\bigm|_{\rm 1-loop}}{\beta} \\
      &=&-\frac{1}{2}\Omega\intq{q}\int\limits_0^1
      d\lambda\;\frac{D_{0E}(q)\Pi_{0E}(q)}{1-\lambda D_{0E}(q)\Pi_{0E}(q)}~, \nonumber
\end{eqnarray}
where now, unlike Eq.~\eqref{eq:A063}, the fraction is truly algebraic.
The above formula coincides (up to the isospin sum)
with the Br\"uckner and Gell--Mann equation for the correlation energy of 
a degenerate electron gas in RPA approximation \cite{FeWa-71-B}. 
Moreover, it is also formally identical to the first--order approximation 
energy of the Modified Loop Expansion derived in Ref.~\cite{We90} for a 
linear $\sigma$ model. It is evident that,
up to the counting operator (namely the integration over $\lambda$),
Eq.~(\ref{eq:A064}) corresponds to a closed loop of the RPA--dressed boson 
propagator.

\subsection{Two--loop correction} 
\label{2l}

By expanding Eq.~\eqref{eq:A035} up to first order 
in the parameter $a$, we find the following formal expression
for the partition function at the next order in the Semiclassical
expansion: 
\begin{eqnarray}
  \label{eq:A065}
      Z(\beta)[J]\bigm|_{2-\rm loops}&=&\Biggl\{1+\frac{\sqrt{a}}{3}
      g^3\, {\mathfrak P}^{(3)}\left(
        \frac{\delta~}{\delta J}\right)^3
      -\frac{a}{4} g^4\, {\mathfrak P}^{(4)}\left(\frac{\delta~}{\delta J}\right)^4\\
      &&+\frac{a}{18}g^6\, {\mathfrak P}^{(3)} \left(
        \frac{\delta~}{\delta J}\right)^3\times{\mathfrak P}^{(3)}\left(
        \frac{\delta~}{\delta J}\right)^3
      \Biggr\} e^{-\frac{1}{2}J\Delta^{\rm RPA}_E J}\Biggm|_{J=0}~. \nonumber
\end{eqnarray}
Here we have defined
\begin{eqnarray}
  \label{eq:A066}
  {\mathfrak P}^{(3)}
  (x_1,x_2,x_3)&=&S_{HE}(x_1-x_2)S_{HE}(x_2-x_3)S_{HE}(x_3-x_1)~, \\
  {\mathfrak P}^{(4)}
  (x_1,x_2,x_3,x_4)&=&S_{HE}(x_1-x_2)S_{HE}(x_2-x_3)S_{HE}(x_3-x_4)
  S_{HE}(x_4-x_1)~, \nonumber
\end{eqnarray}
while ${\mathfrak P}^{(3)}\left(\frac{\delta}{\delta J}\right)^3
\equiv \Tr [S_{HE}\, \frac{\delta~}{\delta J}]^3$ is a short notation for
$$\int dx_1\,dx_2\,dx_3\;{\mathfrak P}^{(3)}(x_1,x_2,x_3)
\frac{\delta~~~}{\delta J(x_1)}\frac{\delta~~~}{\delta J(x_2)}
\frac{\delta~~~}{\delta J(x_3)}~,$$
and an analogous definition holds for the term with four derivatives.

\begin{figure}[ht]
  \begin{center}
    \leavevmode
    \epsfig{file=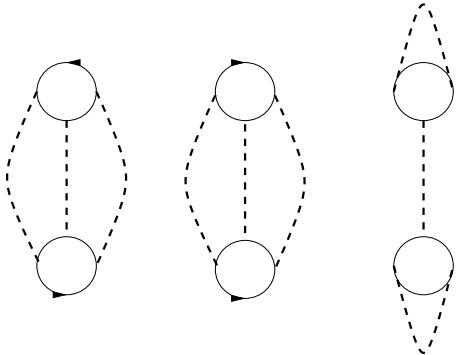,width=6cm,height=4.7cm}
    \caption{Feynman diagrams corresponding, respectively, to the
      contributions $Z^{(3)}_1$, $Z^{(3)}_2$ and $Z^{(3)}_3$ of Eqs.~\eqref{eq:A068}.}
    \label{fig:41}
  \end{center}
\end{figure}

\begin{figure}[ht]
  \begin{center}
    \leavevmode
    \epsfig{file=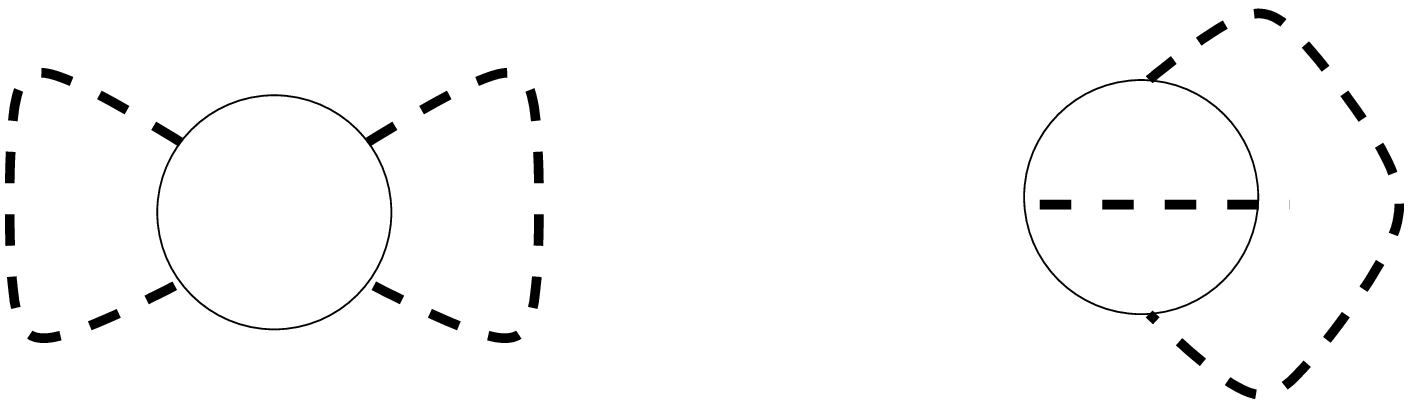,width=4.6cm,height=1.3cm}
    \caption{Feynman diagrams corresponding, respectively, to the
      contributions $Z^{(4)}_1$ and  $Z^{(4)}_2$ of Eq.~\eqref{eq:A069}.}
    \label{fig:42}
  \end{center}
\end{figure}

The term proportional to $\sqrt{a}$ in Eq.~\eqref{eq:A065} vanishes when setting
the source $J$ to 0 and we finally obtain
\begin{equation}
  \label{eq:A067}
  Z(\beta)[J]\bigm|_{2-\rm loops}=1-a\left[-Z^{(3)}_1-Z^{(3)}_2-Z^{(3)}_3
    +Z^{(4)}_1+Z^{(4)}_2\right]~,
\end{equation}
where, analytically,
\begin{eqnarray}
  \label{eq:A068}
  Z^{(3)}_1+Z^{(3)}_2+Z^{(3)}_3&=&\frac{1}{6} g^6
  \int dx_1\,dx_2\,dx_3\;{\mathfrak P}^{(3)}(x_1,x_2,x_3) \\
  &&\times\int dy_1\,dy_2\,dy_3\;{\mathfrak P}^{(3)}(y_1,y_2,y_3)
  \Delta^{\rm RPA}_E(x_1-y_1)\nonumber \\
  &&\times\biggl\{\Delta^{\rm RPA}_E(x_2-y_2)\Delta^{\rm RPA}_E(x_3-y_3)\nonumber \\
  &&+\Delta^{\rm RPA}_E(x_2-y_3)\Delta^{\rm RPA}_E(x_3-y_2)\nonumber \\
  &&+3\Delta^{\rm RPA}_E(x_2-x_3)\Delta^{\rm RPA}_E(y_2-y_3)\biggm\}~, \nonumber
\end{eqnarray}
and
\begin{eqnarray}
  \label{eq:A069}
  Z^{(4)}_1+Z^{(4)}_2&=&\frac{1}{4} g^4
  \int dx_1\,dx_2\,dx_3\,dx_4\;{\mathfrak P}^{(4)}(x_1,x_2,x_3,x_4)\\
  &&\times\biggl\{2\Delta^{\rm RPA}_E(x_1-x_2)\Delta^{\rm RPA}_E(x_3-x_4)\nonumber \\
  &&+\Delta^{\rm RPA}_E(x_1-x_3)\Delta^{\rm RPA}_E(x_2-x_4)\biggm\}~. \nonumber
\end{eqnarray}
In the last two equations, the different contributions to the 
functions $Z^{(3)}_i$ and $Z^{(4)}_i$ appear, in the curly brackets,
according to the order of the lhs terms.
The Feynman diagrams corresponding to these functions
are shown in Figures~\ref{fig:41} and 
\ref{fig:42}. They clearly illustrate how the $Z^{(3)}_i$ and $Z^{(4)}_i$
represent RPA--dressed two--boson--loop contributions. 
Note also that the Feynman diagrams of Figure~\ref{fig:41} (\ref{fig:42})
are the only distinct diagrams containing two boson--loops built up with
three (two) boson propagators. 

At the two--loop level, the ground state energy of nuclear matter is immediately 
obtained from Eq.~\eqref{eq:A067} by expanding 
$\log Z(\beta)[J]\bigm|_{\rm 2-loop}$ at first order in $a$ and then setting $a=1$: 
\begin{eqnarray}
  \label{eq:A070}
   E\bigm|_{2-\rm loops}&\equiv &-
     \lim_{\beta\to\infty}\frac{\log Z(\beta)[J]\bigm|_{\rm 2-loop}}{\beta} \\
     &=&\lim_{\beta\to\infty}
     \frac{1}{\beta}\left[-Z^{(3)}_1-Z^{(3)}_2-Z^{(3)}_3+Z^{(4)}_1
     +Z^{(4)}_2\right]~. \nonumber
\end{eqnarray}
Note that each one of the closed diagrams contributing to
the $Z^{(3)}_i$ and $Z^{(4)}_i$ factorizes a $\beta
\Omega$. Indeed, due to translational invariance, the integrands in
the various terms depend upon one less space--time variable than the number
of space--time integrations. Therefore, the zero temperature limit in
Eq.~\eqref{eq:A070} is finite, as expected, and proportional to $\Omega$.

Finally, from the two--boson--loop approximation for the partition
function 
\begin{equation}
Z(\beta)[J]=Z(\beta)[J]\bigm|_{\rm mean~field} Z(\beta)[J]\bigm|_{1-\rm loop}
Z(\beta)[J]\bigm|_{2-\rm loops}~,
\end{equation}
the ground state energy of nuclear matter is obtained as
\begin{equation}
E=E\bigm|_{\rm mean~field}+E\bigm|_{\rm 1-loop}+E\bigm|_{\rm 2-loops}~,
\end{equation}
with contributions coming from Eqs.~\eqref{eq:A061}, \eqref{eq:A064} and 
\eqref{eq:A070}.
By applying Eqs.~\eqref{eq:010}, \eqref{eq:011} and \eqref{eq:011b},
the binding energy per nucleon, pressure and compressibility of nuclear matter
are thus derived.

\section{Numerical Results}
\label{numerics}

In order to illustrate the convergence properties of the proposed BLE,
in this Section we present a few numerical results obtained in the one--boson--loop 
approximation. Despite the above formal development
has been accomplished for a schematic dynamical model, 
the results discussed in the following correspond to a more realistic
model, QHD-I, which includes the $\omega$ meson beside the scalar--isoscalar one.
In a forthcoming paper we shall discuss the formal modifications introduced by  
QHD-I into the schematic model considered here;
we shall also systematically discuss the predictions for the
static properties of nuclear matter
of the $\sigma$-$\omega$ model up to the two--boson--loop order.

\begin{table}
\begin{center}
\caption{Coupling constants of the QHD-I model at the mean field and one--boson--loop
approximations.}
\label{parametri}
\begin{tabular}{l| c c}\hline
\mc {1}{c|}{} &
\mc {1}{c}{$g_\sigma$} &
\mc {1}{c}{$g_\omega$} \\ \hline
Mean Field & 11.09 & 13.81 \\
One--Boson--Loop   & 13.00 & 15.45 \\ \hline
\end{tabular}
\end{center}
\end{table}
In our calculations we have adopted the following meson masses:
$m_\sigma= 550$ MeV and $m_\omega=783$ MeV. First we have investigated
the properties of symmetric nuclear matter obtained at the
one--boson--loop level by
adopting the values of the coupling constants $g_\sigma$ and $g_\omega$ which 
allow us to reproduce the empirical saturation point of nuclear matter
($B.E./A=-16.0$ MeV and $k_F=1.3$ fm$^{-1}$)
at the mean field level (see Table~\ref{parametri}).
For Fermi momenta larger than $0.2$ fm$^{-1}$ 
the one--boson--loop binding energy turned out to be repulsive, with 
$B.E./A=+18.6$ MeV for $k_F=1.3$ fm$^{-1}$. This is mainly due
to a strongly repulsive Fock term for the $\sigma$ which is not compensated 
by the attractive Fock contribution for the $\omega$. 
By adopting the same criteria for convergence established by
Ref.~\cite{Furnstahl:1989wp} for the Loop Expansion and mentioned 
in the Introduction, we can conclude that 
the above large changes foreclose a ``strong'' convergence of the BLE. 
We then realise that even a non--perturbative expansion in the ``large'' couplings
$g_\sigma$ and $g_\omega$ such as the BLE turns out not to be convergent in the ``strong'' 
sense. We remind the reader that the same finding was attained for the Loop Expansion
\cite{Furnstahl:1989wp}, which, on the contrary, is perturbative in the couplings.

\begin{figure}[t]
  \begin{center}
    \epsfig{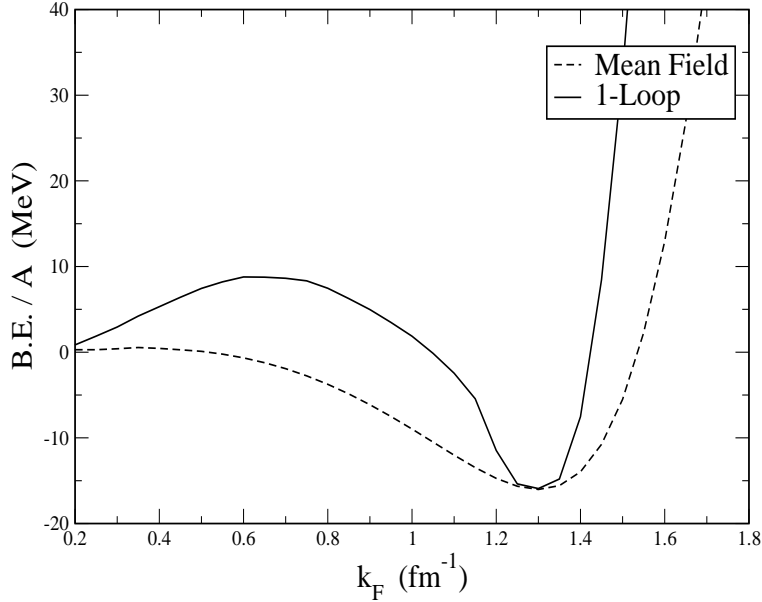}
    \caption{Predictions of a $\sigma$-$\omega$ model for the
binding energy per nucleon of symmetric nuclear matter
obtained in the mean field and one--boson--loop approximations.
The adopted coupling constants are given in Table~\ref{parametri}.}
    \label{numeri}
  \end{center}
\end{figure}

It thus becomes important to investigate on a possible ``weak'' convergence
of the BLE. To establish the validity of such a property, we have to check if
it is possible to reproduce 
the empirical saturation of nuclear matter in the one--boson--loop approximation
by a ``reasonable'' refit of the parameters. 
In Figure~\ref{numeri} we plot the binding energy per nucleon of symmetric
nuclear matter which we obtained in the one--boson--loop approximations by adopting
the couplings of Table~\ref{parametri}.
The well known mean field predictions are displayed as well to facilitate
the comparison with the result of the BLE at first order.
Since the obtained refit modifies the
couplings by a small amount (of about 15\%), we can conclude that,
contrary to the Loop Expansion of Ref.~\cite{Furnstahl:1989wp}, the BLE
is weakly convergent. Note that such an adjustment of the couplings practically
maintains unaltered the underlying non--relativistic nucleon--nucleon interaction 
of the $\sigma$-$\omega$ model.

We can therefore conclude that the scheme provided by
the BLE proves to be a good candidate for performing systematic and
reliable calculations, based on QHD models, beyond the tree--level approximation.

\section{Conclusions and Perspectives}
\label{sec:6X}

In this paper we have proposed a functional integral approach
to study the static properties of nuclear matter within a fully relativistic 
scheme, keeping in mind, even if not explicitly considered, the QHD model 
developed by Serot and Walecka. First, a bosonic effective
action has been constructed for a system of interacting nucleons and $\sigma$ 
mesons. Special attention has been devoted to the formulation of a
renormalization procedure for the introduced bosonic effective action.
We have then evaluated the resulting partition function
through a Semiclassical expansion
around the solution given by the stationary phase approximation: 
this allowed us to express the result in terms of a boson loop expansion.
Hartree--dressed nucleons and RPA--dressed bosons have
revealed themselves to be the basic elements of this BLE.
The suggested scheme has been analyzed
at the mean field level and at the one-- and two--boson--loop orders.
We have presented a thorough exposition of the formal derivation
of the binding energy per
nucleon, the pressure and the compressibility of nuclear matter
up to the second order in the BLE.

Numerical results for the nuclear matter binding energy at the
one--boson--loop level have been presented for a realistic, dynamical model including
the $\omega$ meson in addition to the $\sigma$.
The convergence properties of the BLE have been discussed in this context.
A systematic calculation of the static properties of nuclear matter 
up to the two--boson--loop order and in the $\sigma$-$\omega$ model will
be the object of a forthcoming paper, where
we shall also deal with the formal derivation of the higher orders of 
the BLE. 
The inclusion of the $\omega$ meson
is not a prohibitive task from the point of view of the diagrammatic content,
since the basic topology of the boson loops (see Figures~\ref{fig:41} and
\ref{fig:42}) is common to any meson field which can contribute to the dynamics.
However, since the $\omega^0$ component is coupled, in infinite nuclear matter, to 
the $\sigma$ meson (due to the breaking of the Lorentz invariance induced by the 
infinite medium) \cite{Du-03}, the RPA equations, which deeply intervene in the
corrections beyond the mean field, are much more involved that those 
presented here.
                
Further, we note that the implementation of the isovector mesons $\rho$ and
$\delta\, [f_0(980)]$ in our framework does not introduce additional difficulties 
from a technical point of view. Up to now, however, they have
been handled only at the mean field level \cite{Li-al-02}.

It is worth noticing that at the mean field level no pseudoscalar and pseudovector
mesons can be coupled to the nucleon density or current, since the
average values of pseudoscalar and pseudovector quantities vanish. However,
the situation is completely different when boson loops are involved.
It is indeed known
that the two--pion exchange diagrams, with the possible simultaneous excitation
of one or two intermediate nucleons to a $\Delta$, provide a large amount of the nuclear
binding. We shall explore this mechanism in the future, but we reasonably
expect that a drastic change in the underlying dynamics of QHD will occur
and that the gap between QHD and Bonn potential models will be further, and 
significantly, reduced.

\appendix
\section*{Appendix}

\label{sec:appA}

In the regularization scheme expressed by Eq.~\eqref{eq:H270} the explicit
expression of the scalar density,
$\rho_s(m)=\rho_{s\, v}(m)+\rho_{s M}(m)$, is given by:
\begin{eqnarray}
  \label{eq:M272}
  \rho_{s\, v}(m)&=&\frac{m\Lambda^4}{2\pi^2(m^2-\Lambda^2)}\left\{1
    +\frac{m^2}{m^2-\Lambda^2}\log\frac{\Lambda^2}{m^2}\right\}~, \nonumber \\
  \rho_{s M}(m)&=&\frac{m\Lambda^4}{\pi^2(m^2-\Lambda^2)^2}
  \left\{k_FE_F-m^2\log\frac{k_F+E_F}{m}\right\}~, \nonumber
\end{eqnarray}
with the limiting behavior
\begin{equation}
  \label{eq:H906}
  \rho_{s\, v}(m) \mathop{\longrightarrow}_{\Lambda\to \infty}
-\frac{m}{2\pi^2}\left(\Lambda^2-2m^2
    \log\frac{\Lambda}{m}+m^2\right)~. \nonumber
\end{equation}

The quantity $\Pi^{(d)}_v(q^2)$
introduced in Eqs.~\eqref{eq:H5011} and \eqref{eq:H905} reads
\begin{eqnarray}
  &&\Pi^{(d)}_v(q^2)=\frac{\left(4
      m^2-q^2\right)^{3/2} 
     \arctan\left(\frac{\displaystyle q}{\displaystyle \sqrt{4
          m^2-q^2}}\right) \Lambda^4}{2 \pi ^2 q (m-\Lambda
    )^2 (m+\Lambda)^2} \nonumber \\
  &&-\bigl(-m^6+11
  q^2 m^4+3 \Lambda ^2 m^4-7
  q^4 m^2-3 \Lambda ^4 m^2+6
  q^2 \Lambda ^2\nonumber \\
  &&+m^2+q^6+\Lambda ^6-q^2
  \Lambda ^4-q^4 \Lambda^2\bigr)\nonumber \\
  &&\times \frac{\Lambda^4}{4 \pi ^2 q^2 (m^2-\Lambda^2
    )^2 \sqrt{m-q-\Lambda }
    \sqrt{m+q-\Lambda }
    \sqrt{m-q+\Lambda }
    \sqrt{m+q+\Lambda
      }}\nonumber \\
  &&\times \Biggl[{\rm arctanh}
    \left(\frac{m^2-q^2-\Lambda ^2}{\sqrt{m-q-\Lambda
          } \sqrt{m+q-\Lambda }
        \sqrt{m-q+\Lambda }
        \sqrt{m+q+\Lambda
          }}\right)\nonumber \\
    &&-{\rm arctanh}
    \left(\frac{m^2+q^2-\Lambda ^2}{\sqrt{m-q-\Lambda
          } \sqrt{m+q-\Lambda }
        \sqrt{m-q+\Lambda }
        \sqrt{m+q+\Lambda
          }}\right)\Biggr]\nonumber \\ 
   &&-\frac{\left(m^4-6 q^2
   m^2-2 \Lambda ^2
   m^2+q^4+\Lambda ^4\right)
   \log \left(\frac{\displaystyle m}{\displaystyle \Lambda
   }\right) \Lambda ^4}{4 \pi
   ^2 q^2 (m^2-\Lambda^2 )^2} 
   -\frac{\Lambda ^4}{4\pi ^2 \left(m^2-\Lambda^2\right)}~. \nonumber
\end{eqnarray}

In the limit $\Lambda\to\infty$ and within a hard normalization scheme
we find
\begin{eqnarray}
  \Pi^{(d)}_v(m_\sigma^2)&=&\frac{1}{2\pi^2}   \Biggl\{\Lambda^2-
  (6m^2-m^2_{\sigma})\log\frac{\Lambda}{m}\nonumber\\
  &&+\frac{4m^2+m^2_{\sigma}}{4}
  +\frac{(4m^2-m^2_{\sigma})^\frac{3}{2}}{m_{\sigma}}
  \arctan\frac{m_{\sigma}}{\sqrt{4m^2-m^2_{\sigma}}}\Biggr\}~,\nonumber
\end{eqnarray}
and
\begin{eqnarray}
  \frac{d\Pi^{(d)}_v(q^2)}{dq^2}\Bigm|_{q^2=m_\sigma^2}
  &=&\frac{1}{2\pi^2}\Biggl\{-\frac{1}{4}+\frac{2m^2}{m_{\sigma}^2}\nonumber \\
    &&-\frac{\sqrt{4m^2-m_{\sigma}^2}(2m^2+m_{\sigma}^2)}{m_{\sigma}^3}
    \arctan\frac{m_{\sigma}}{\sqrt{4m^2-m_{\sigma}^2}}
    +\log\frac{\Lambda}{m}\Biggr\}~. \nonumber
\end{eqnarray}
On the contrary, by choosing the renormalization point at $q^2=0$ we find
\begin{equation}
  \label{eq:H303}
  \begin{split}
    \Pi^{(d)}_v(0)&=\frac{\Lambda^4}{2\pi^2}
    \left\{\frac{3m^2+\Lambda^2}{(\Lambda^2-m^2)^2}
      -2m^2\frac{3\Lambda^2+m^2}{(\Lambda^2-m^2)^3}\log\frac{\Lambda}{m}
      \right\}\\
      &\mathop{\longrightarrow}_{\Lambda\to \infty} 
        \frac{1}{2\pi^2}\left\{\Lambda^2-6m^2\log\frac{\Lambda}{m}
        +5m^2\right\}~, \nonumber
  \end{split}
\end{equation}
and
\begin{equation}
  \label{eq:H304}
  \begin{split}
    \frac{d\Pi^{(d)}_v(q^2)}{dq^2}\Bigm|_{q^2=0}&=
    \Biggl\{\frac{-\Lambda^4(13\Lambda^4-68m^2\Lambda^2-41m^4)}{24\pi^2
      (\Lambda^2-m^2)^4}\\
    &+\frac{\Lambda^4(\Lambda^6-3m^2\Lambda^4
      -12m^4\Lambda^2-2m^6}{2\pi^2(\Lambda^2-m^2)^5}
    \log\frac{\Lambda}{m}\Biggr\}\\
      &\mathop{\longrightarrow}_{\Lambda\to \infty}
       -\frac{13}{24\pi^2}+\frac{1}{2\pi^2}\log\frac{\Lambda}{m}~. \nonumber
  \end{split}
\end{equation}

For the $\sigma\sigma\sigma$ vertex function of Eq.~(\ref{eq:H910}) we have
\begin{equation}
  \begin{split}
    \widetilde\Gamma^{(e)}&=\frac{2 m \Lambda ^4
      \left(m^2+2 \Lambda
        ^2\right)}{\pi ^2
      (m-\Lambda )^3 (m+\Lambda
      )^3}
    +\frac{m \Lambda ^4
      \left(m^4+8 \Lambda ^2
        m^2+3 \Lambda ^4\right)
      }{\pi ^2 (m-\Lambda )^4
      (m+\Lambda )^4}\log\frac{\Lambda}{m}\\
    &\mathop{\longrightarrow}_{\Lambda\to \infty}
     -\frac{4m}{\pi^2}+\frac{3m}{\pi^2}\log\frac{\Lambda}{m} \nonumber
  \end{split}
\end{equation}
and for the $\sigma\sigma\sigma\sigma$ self--coupling of
Eq.~(\ref{eq:H9100}):
\begin{equation}
  \begin{split}
    \widetilde\Gamma^{(f)}&=-\frac{\left(7 m^4+34 \Lambda
        ^2 m^2+7 \Lambda ^4\right)
      \Lambda ^4}{3 \pi ^2
      (\Lambda -m)^4 (m+\Lambda
      )^4}
     +\frac{\left(m^2+\Lambda ^2\right) \left(m^4+14
        \Lambda ^2 m^2+\Lambda
        ^4\right)  \Lambda ^4}{\pi
      ^2 (\Lambda -m)^5
      (m+\Lambda )^5}\log\frac{\Lambda}{m}\\
    \nonumber &\mathop{\longrightarrow}_{\Lambda\to \infty}
    -\frac{7}{3\pi^2}+\frac{1}{\pi^2}\log\frac{\Lambda}{m}~.
  \end{split}
\end{equation}

\end{document}